\documentclass{article}

\usepackage{PRIMEarxiv}

\usepackage[utf8]{inputenc} % allow utf-8 input
\ifdefined\XeTeXversion
  \usepackage{fontspec}
\else
  \usepackage[T1]{fontenc}
\fi
\usepackage{hyperref}       % hyperlinks
\usepackage{url}            % simple URL typesetting
\usepackage{booktabs}       % professional-quality tables
\usepackage{amsfonts}       % blackboard math symbols
\usepackage{nicefrac}       % compact symbols for 1/2, etc.
\usepackage{microtype}      % microtypography
\usepackage{lipsum}
\usepackage{fancyhdr}       % header
\usepackage{graphicx}       % graphics
\usepackage[labelfont=bf,labelsep=period,singlelinecheck=false]{caption}
\usepackage{amsmath}
\usepackage{multirow}
\usepackage{multicol}
\usepackage{array}
\usepackage[section]{placeins}
\usepackage{tabularx}       % tables fitted to the text width
\usepackage{pifont}         % \ding (cross mark)
\newcommand{\cmark}{\ding{51}}
\newcommand{\xmark}{\ding{55}}
\graphicspath{{media/}}     % organize your images and other figures under media/ folder

\ifdefined\pdflinkmargin
\else
  \ifdefined\XeTeXLinkMargin
  \fi
\fi

\title{{\upshape Reference Manual of \textit{MEDiml}}}

\author{
  \textbf{Mahdi Ait Lhaj Loutfi$^1$, Simon Provost$^1$, Élodie Gillard$^2$, Moustafa Amine Bezzahi$^2$, Alexander Chum$^1$, Ben Coull-Neveu$^1$, Juan Duran$^1$, Mariem Kallel$^2$, Hakima Laribi$^2$, Oliver Lefebvre$^2$, Nicolas Longchamps$^2$, Thanh-Tri Nguyen$^1$, Steven Ravary$^1$, Ouael Nedjem Eddine Sahbi$^2$, Jérémy Dana$^{3}$, Martin Vallières$^{1,2*}$}\\
  \vskip 0.2in \center{}   \vskip 0.2in
  $^1$Medical Physics Unit, Department of Oncology, McGill University, Montreal, Canada
  \\$^2$Department of Computer Science, Faculty of Science, University of Sherbrooke, Sherbrooke, Canada
  \\$^3$Department of Radiology, Centre de Recherche du Centre Hospitalier de l'Université de Montréal, Montréal, Canada
  \\$^{*}$Corresponding author: \texttt{martin.vallieres@mcgill.ca}
}

\begin{document}
\maketitle

\vspace{-0.5cm}

\begin{abstract}

Radiomics is the high-throughput extraction of quantitative features (e.g., tumor shape, intensity distributions and texture) from medical images to characterize tissues and uncover disease patterns invisible to the human eye. These features can be evaluated using statistical and machine learning methods as candidate imaging biomarkers for cancer diagnosis, prognosis and classification. However, extracting thousands of candidate features makes it difficult to identify which measurements are most informative for a given clinical task, while implementing an end-to-end workflow can require programming expertise that limits its accessibility to clinical researchers. Hence, we propose \textit{MEDiml} (MEDical Image analysis with Machine Learning), an open-source radiomics research platform compliant with international standards, which connects feature extraction with predictive modeling and interpretation through complementary graphical and programmatic interfaces.\\

\textit{MEDiml} combines a Python package for radiomic feature extraction and predictive modeling (\textit{MEDiml-pkg}) with a graphical desktop application (\textit{MEDiml-app}) that enables researchers to configure and conduct experiments without writing code. The two interfaces use the same computational implementation, and code generation exports graphically configured workflows as executable Python code, facilitating inspection, reproducibility, extension, and collaboration between clinical researchers and computational scientists.\\

A typical \textit{MEDiml} workflow includes loading medical images and segmentation masks defining the regions of interest, preprocessing and filtering, extracting radiomic features, reducing the feature set, training and evaluating predictive models, and interpreting model behavior using feature importance computations. Feature sets can be grouped into predefined levels of radiomic complexity, including morphological, intensity-based, textural, and filter-based features to support comparisons of predictive performance across these levels and evaluate whether increasingly complex feature groups provide additional predictive value.\\

This manuscript serves as the official \textit{MEDiml} \textbf{Reference Manual}. It introduces the relevant radiomics concepts, documents the analysis workflow and the \textit{MEDiml-pkg} and \textit{MEDiml-app} components, and describes the software architecture and installation procedures. It also compares \textit{MEDiml} with existing radiomics tools and demonstrates its complexity-driven workflow through use cases involving five oncological imaging cohorts comprising 2,104 scans. Links to the software distributions, source code, documentation, tutorials, and community-support resources are available at \href{https://www.mediml.app}{mediml.app}. 

\end{abstract}

% keywords can be removed
\keywords{software engineering, open-source, radiomics, clinical research, image processing, machine learning, medical image analysis}

\clearpage
\setcounter{tocdepth}{2}
\tableofcontents
\clearpage

\section{Introduction}
\label{sec:introduction}
\begin{figure}[!htbp]
  \centering
  \includegraphics[scale=0.18]{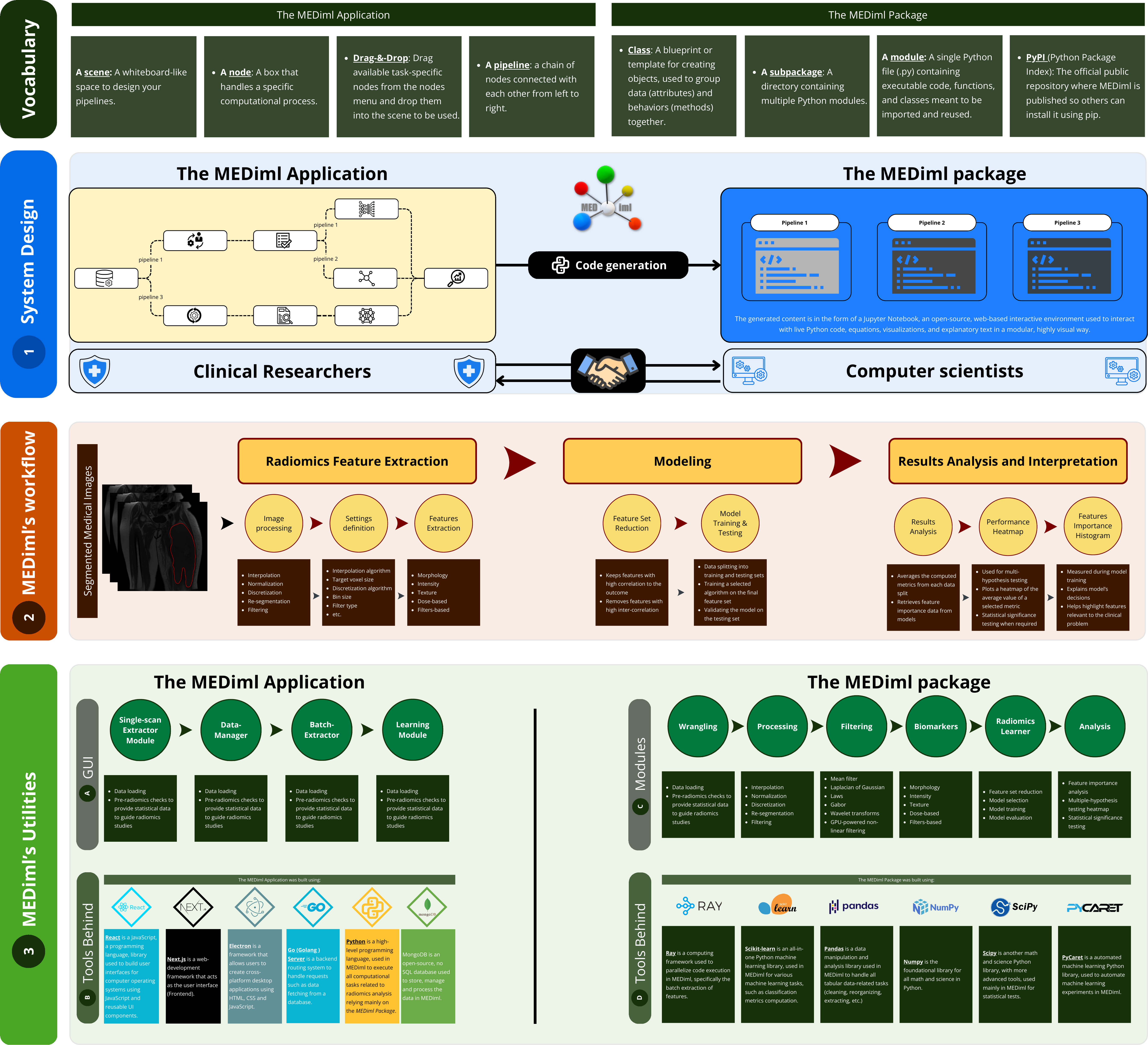}
  \caption{\textbf{Overview of the \textit{MEDiml} application and Python package}. Block 1 shows the system design of both components and the code-generation feature that connects graphical workflows with Python code. Block 2 presents the analysis workflow, which starts with feature extraction from processed medical images, and continues with feature set reduction, model training, and results analysis. Block 3 shows the modules available through each component and the technologies on which they depend.}
  \label{fig:introductory}
\end{figure}

Medical imaging is a cornerstone of personalized medicine, as it provides a non-invasive window into the unique phenotypic characteristics of volumes of interest. Radiomics is defined as the high-throughput extraction of quantitative features from images to enable characterization of tissues \cite{lambin_radiomics_2012}. It has the potential to help with diagnosis, assess prognosis, and provide clinicians with new tools to enable more precise personalized treatment \cite{images_more_data}. Radiomics features extend our ability to discern subtle nuances in medical imaging characteristics, providing data for predictive modeling approaches to enhance personalized treatment \cite{lambin_radiomics_2017}.

Handcrafted radiomic analysis encompasses various feature categories, each potentially related to different aspects of a tissue phenotype. These categories include morphological features, which describe the shape and size of analyzed regions; intensity features, which capture characteristics related to pixel intensity distributions; and texture features, which quantify spatial intensity patterns. These features can be extracted from the original image intensities, such as Hounsfield Units (HU) in X-ray computed tomography (CT) or arbitrary intensities in magnetic resonance imaging (MRI), but also from image intensities previously modified by filters that highlight various structures and patterns \cite{ibsi_2}. For the sake of readability, the term radiomics in this manual refers specifically to handcrafted radiomics, unless otherwise specified. Moreover, the Image Biomarker Standardization Initiative (IBSI), an independent international collaboration, established a nomenclature and definitions for commonly used radiomic features, developed a general image processing workflow for calculating radiomic features, and provided datasets and associated reference values for verifying and calibrating software implementations \cite{ibsi_1}.

Despite radiomics' expanding role in research, and despite the standardization efforts, few tools have reached clinical practice. Most tools require programming experience, leaving clinical researchers without access to user-friendly research tools with minimal functional requirements. At the same time, doing radiomics analysis requires bridging two disciplines: clinical research, which handles the medical questions, and computer science, which handles the computational issues. Most existing radiomic tools serve only one discipline, not both.

In this reference manual, we present MEDical Image analysis with Machine Learning (\textit{MEDiml}) \textit{v1.0}, an IBSI-compliant \cite{ibsi_1, ibsi_2} open-source platform designed to improve the accessibility of radiomics for clinical researchers and facilitate collaboration between multidisciplinary teams. Here, \textit{MEDiml} denotes the complete project, including its scientific approach, software components and supporting resources. Figure \ref{fig:introductory} provides an overview of \textit{MEDiml} for reference throughout the manual.

The contributions of \textit{MEDiml} can be summarized as follows: First, it provides a complexity-driven workflow that compares models built from progressively more complex radiomic feature sets, so that researchers can identify a compact feature set suited to a clinical question \cite{loutfi_complexity_2024}. Second, its Python package (\textit{MEDiml-pkg}) and graphical application (\textit{MEDiml-app}) share the same computational framework, so that code generation exposes visually configured workflows as Python code that researchers can inspect and extend. The two interfaces therefore allow researchers with different programming experience to use the same analysis implementation. Third, \textit{MEDiml} provides the Explanation Tree, a visualization of model-specific feature importance proposed in this manual that organizes feature contributions hierarchically, complementing statistical model comparisons and feature-importance histograms.

The remainder of this manual is structured as follows: Section~\ref{sec:resources} provides the software distributions, source-code repositories, documentation, tutorials and community resources. Section \ref{sec:background_radiomics} provides the radiomics background. Section \ref{sec:mediml_workflow} describes the \textit{MEDiml} radiomics pipeline, including feature extraction, feature set reduction, model training and evaluation, and result interpretation. This section also presents the concept of complexity levels, which forms the methodological foundation of the platform and guides the analysis workflow. Subsection \ref{sec:install} explains how to install \textit{MEDiml}. Section \ref{sec:mediml_pkg} describes the Python package and Section \ref{sec:mediml_app} describes the zero-code graphical user interface (GUI), with both sections providing user-oriented documentation. Section \ref{sec:sys_design} describes the software architecture, including the code-generation feature that translates visually designed workflows into executable Python code. In Section \ref{sec:mediml_vs_rest}, we compare \textit{MEDiml} to existing radiomics software tools. Section \ref{sec:use_cases} presents scientific use-cases evaluating \textit{MEDiml} across oncological imaging datasets spanning different clinical problems and imaging modalities. Section \ref{sec:discussion} discusses the software, the use-cases, limitations and future work. Finally, Section \ref{sec:conclusion} provides concluding remarks.

\section{Resources}
\label{sec:resources}
The official \textit{MEDiml} website, \url{https://www.mediml.app}, is the main entry point for the platform, providing an overview of its capabilities and links to software downloads, documentation and tutorials. The resources below distinguish the Python package (\textit{MEDiml-pkg}) from the desktop application (\textit{MEDiml-app}), so that the reader can easily locate the material appropriate to the chosen interface.

\subsection{Python Package}
\label{sec:resources_pkg}
\paragraph{Distribution.}
The Python Package Index (PyPI), \url{https://pypi.org/project/mediml/}, provides the installable Python package, release history and distribution files. Installation, including the selection of a specific package version to reproduce an existing analysis, is described further in Subsection~\ref{sec:install_pkg}.

\paragraph{Documentation.}
The Python package documentation, available at \url{https://mediml.readthedocs.io/en/latest/}, covers installation, input-data organization, configuration files and the application programming interface (API).

\paragraph{Source Code and Issue Tracking.}
The Python package repository, \url{https://github.com/MEDomicsLab/MEDiml-pkg}, provides the source code, development history and contribution material. Its issue tracker, \url{https://github.com/MEDomicsLab/MEDiml-pkg/issues}, can be used to report reproducible problems or request new package features.

\subsection{Desktop Application}
\label{sec:resources_app}
\paragraph{Distribution.}
The download section of the official website, \url{https://www.mediml.app/\#download}, provides desktop application installation files for \textit{Windows}, \textit{macOS} and \textit{Linux}. Installation and initial environment setup are described further in Subsection~\ref{sec:install_app}.

\paragraph{Documentation.}
The desktop application documentation, available at \url{https://medomicslab.gitbook.io/mediml-app-docs}, covers installation, graphical workflow construction and the application modules. It also provides troubleshooting and contribution guidance, complementing the interface descriptions in Section~\ref{sec:mediml_app}.

\paragraph{End-to-End Demonstration} An \href{https://medomicslab.gitbook.io/mediml-app-docs/soft-tissue-sarcoma-poc}{end-to-end demonstration} of \textit{MEDiml} on real-world data of soft-tissue sarcoma is also included~\cite{radiomics_10k_6}.

\paragraph{Video Tutorials.}
The YouTube tutorial playlist, \url{https://www.youtube.com/playlist?list=PLEPy2VhC4-D5Eg-UxRyTtmUZRh-D5m_Ru}, demonstrates the desktop application workflow, from image loading and feature extraction to model training and results analysis.

\paragraph{Source Code and Issue Tracking.}
The desktop application repository, \url{https://github.com/MEDomicsLab/MEDiml-app}, provides the source code, development history and contribution material. Its issue tracker, \url{https://github.com/MEDomicsLab/MEDiml-app/issues}, can be used to report reproducible problems or request new application features.

\subsection{Community and Development}
\label{sec:resources_community}
The community Discord server, in the \textit{MEDiml} channel, \url{https://discord.gg/GjMW72fE9T}, provides a shared discussion space for users and developers. For problems that require a reproducible report, the issue tracker of the affected component provides a record that can be followed during development.

Finally, \textit{MEDiml} is developed and maintained within MEDomicsLab, whose website, \url{https://www.medomicslab.com}, provides information about the research group and its research activities.

\section{Background: Radiomics}
\label{sec:background_radiomics}
As mentioned, in Section \ref{sec:introduction}, radiomics is defined as the high-throughput extraction of quantitative features from images to enable characterization of tissues. These features are used to train machine learning models to diagnose, classify or predict clinical outcomes. These features depend on how the images are processed and which measurements are calculated within the region of interest. The IBSI provided definitions and reference values for 173 features. The content presented in this section is based on the IBSI manuscripts \cite{ibsi_1, ibsi_2}. The remainder of this section is structured as follows: Subsection \ref{sec:img_processing_radiomics} describes the image-processing operations, and Subsection \ref{sec:Image_Features} describes the radiomic feature families.

\subsection{Image Processing}
\label{sec:img_processing_radiomics}
Image processing refers to performing operations on digital images to improve image quality or extract useful information. It takes an image as an input and outputs a modified and improved image, while the applied operations are selected according to the input image and the desired output. For instance, denoising MR scans prior to automatic segmentation can facilitate the detection of objects within the image. As such, all image processing related tasks are handled by the processing subpackage in \textit{MEDiml}.

\subsubsection{Segmentation}
\label{sec:segmentation}
Radiomics feature extraction relies on the segmentation of medical images, where a Region of Interest (ROI) is delineated to define an anatomical structure for the analysis, such as a brain tumor. \textit{MEDiml} uses its processing subpackage to load and process segmentation masks, which identify the voxels belonging to the ROI. These masks are often represented as binary, where a value of 1 denotes a voxel belonging to the ROI and a value of 0 denotes a voxel outside the ROI. The mask is used throughout the analysis to restrict feature extraction to the relevant anatomical region.

\subsubsection{Interpolation}
\label{sec:interpolation}
Many image features are sensitive to changes in voxel size. Maintaining consistent isotropic voxel spacing across different measurements and devices is therefore important for reproducibility, and interpolation to isotropic voxel spacing is required. While numerous interpolation algorithms and resampling strategies have been proposed, there is currently no clear consensus regarding the advantage of a particular method or whether upsampling and downsampling approaches should be preferred. The IBSI guidelines \cite{ibsi_1} recommend trilinear interpolation as a conservative default option. Accordingly, \textit{MEDiml} supports multiple interpolation methods, allowing users to select the approach most appropriate for their dataset and study design.

\subsubsection{Re-segmentation}
\label{sec:re_segmentation}
Re-segmentation is the process of updating the ROI to exclude a range of voxels from the previously segmented ROI. For example, excluding bone and air voxels from CT images.

\subsubsection{ROI extraction}
\label{sec:roi_extraction}
Certain features require the ROI mask to be isolated from the imaging volume. Therefore, the voxels outside the ROI are assigned a placeholder, for example: \textit{NaN}, which stands for “Not a Number”, marking them as excluded from subsequent calculations. Voxels within the ROI mask retain their original intensity.

\subsubsection{Discretization}
\label{sec:discretization}
Discretization, also referred to as quantization, is a key step to make calculation of texture features tractable. Two approaches are used: Fixed Bin Number, where intensities are discretized to a user-defined bin number, and Fixed Bin Width, where the bins’ width is set by the user, i.e., each intensity interval with the pre-selected bin width is assigned a unique bin. Other methods exist, such as the intensity histogram equalization \cite{radiomics_10k_6}. \textit{MEDiml} implements all the aforementioned methods.

\subsection{Image Features}
\label{sec:Image_Features}
Radiomics analysis enables the extraction of a large number of quantitative features, which can be divided into sets of families, defined according to the aspects they describe in the ROI. In the \textit{MEDiml} application, feature extraction is done through the \textit{Single-Scan Extraction} and the \textit{BatchExtractor} modules. For the \textit{MEDiml} Package, the biomarkers subpackage handles all the extraction.

\subsubsection{Morphological features}
\label{sec:Morphological_features}
Morphological features characterize the geometric properties of the ROI. In accordance with IBSI definitions \cite{ibsi_1}, a total of 29 three-dimensional morphological features can be computed from the segmented volume. Their units depend on the measurement (e.g., millimeters for length), with squared or cubed units for area or volume, respectively, while some shape measures are dimensionless. By default, \textit{MEDiml} computes 27 morphological features and excludes Moran's $I$ Index and Geary's $C$ Measure, as they are computationally expensive to extract. Users may optionally enable the calculation of these measures by explicitly specifying the corresponding extraction parameters.

\subsubsection{Intensity features}
\label{sec:Intensity_features}
Intensity features describe the distribution of voxel intensities within the ROI, without considering the spatial arrangement of the voxels. In accordance with IBSI definitions \cite{ibsi_1}, they are divided into four families: local intensity features, intensity-based statistical features, intensity histogram features, and intensity-volume histogram features. Local intensity and statistical features are calculated directly from the voxel intensities, whereas intensity histogram and intensity-volume histogram features are calculated from discretized intensities (see Subsection \ref{sec:discretization}). Each family is described below.

\paragraph{\textit{Local Intensity features}.}
\label{sec:loc_intensity_features}
Local intensity features include two measures that are calculated using the intensities of voxels within a defined neighborhood around a central voxel. Unlike many other feature sets, local intensity features do not rely solely on intensities within the ROI. While only voxels from the ROI intensity map can be used as central voxels, the local neighborhood includes all voxels, regardless of whether they are located within an ROI.

\paragraph{\textit{Statistical features}.}
\label{sec:stat_features}
A total of 18 characteristics quantify the distribution of voxel intensity values within an ROI. They include statistical measures such as the mean, median, standard deviation, and variance, among others. They capture essential information about the variation in pixel intensity within the ROI, thereby providing insights into tissue homogeneity.

\paragraph{\textit{Intensity Histogram features}.}
\label{sec:int_hist_features}
These 23 characteristics quantify the distribution of voxel intensity values within an ROI using statistics derived from the intensity histogram. After discretizing the original intensity distribution into intensity intervals, that is, with each interval representing a range of intensity values, the histogram is constructed by counting the number of voxels in each interval of the discretized intensity.

\paragraph{\textit{Intensity Volume Histogram features}.}
\label{sec:int_vol_hist_features}
These five characteristics are based on the intensity-volume histogram and are used to quantify the distribution of intensity values in relation to the volume of an ROI. The intensity values in the ROI are discretized into a user-defined number of gray levels. For each discretized intensity level, the fraction of the total volume of the ROI with an intensity greater than or equal to that level is calculated. This creates a cumulative relationship between intensity and volume. Finally, a cumulative histogram of the volumes corresponding to each intensity level is constructed. The features extracted from this histogram include measures such as the volume fraction at the 10th percentile of intensities, as well as the volumes at specific intensities, allowing for a detailed assessment of the distribution of intensities as a function of volume within the segmented volume.

\subsubsection{Texture features}
\label{sec:Texture_features}
Texture features capture local variations in intensity or color within an image. They provide information about the spatial distribution of gray levels or voxel values, which can be useful for distinguishing between different tissues or for detecting abnormal structures. Texture features may include measures such as variance, homogeneity, roughness, entropy. Image intensities are generally discretized before calculation of texture features.
In accordance with IBSI definitions \cite{ibsi_1}, texture features are calculated from six matrix families, each summarizing a different type of spatial relationship between discretized gray levels:
\begin{itemize}
    \item Gray-Level Co-occurrence Matrix (GLCM) \cite{sebastian_glcm_2012}: counts how often pairs of gray levels occur in neighboring voxels along a given direction. 25 features are defined in this family;
    \item Gray-Level Run Length Matrix (GLRLM): counts runs, i.e., consecutive voxels with the same gray level along a given direction, according to their gray level and length;
    \item Gray-Level Size Zone Matrix (GLSZM): counts zones, i.e., groups of connected voxels sharing the same gray level, according to their gray level and size;
    \item Gray-Level Distance Zone Matrix (GLDZM): counts zones according to their gray level and their distance to the edge of the ROI;
    \item Neighborhood Gray Tone Difference Matrix (NGTDM): sums the differences between the gray level of each voxel and the average gray level of its neighborhood;
    \item Neighboring Gray Level Dependence Matrix (NGLDM): counts how many neighboring voxels depend on, i.e., have a gray level similar to, each central voxel.
\end{itemize}

\subsubsection{Filter-based features}
\label{sec:Filters-based_features}
This set represents the image features calculated after applying a filter on input images. Applying a filter alters the intensities of the image voxels, which can change their original physical meaning. Consequently, only features based on the intensity histogram and those related to texture are calculated after filtering. These features capture important information about the distribution of gray levels in the image as well as the textural patterns present, which is often more relevant in the context of post-filtering radiomic analysis.

\section{\textit{MEDiml} radiomics workflow}
\label{sec:mediml_workflow}
The background distinguishes the operations used to prepare an image from the features extracted from it. In an analysis, these choices form part of a wider workflow that also includes model training and interpretation. \textit{MEDiml} connects these stages so that feature sets downstream can be compared according to their radiomic complexity.

Given the different image preprocessing steps, such as interpolation and intensity range definition, that may be performed prior to feature extraction, some radiomic features, such as the mean of gray level intensities, can be simple quantities to calculate, while others, such as texture features, are more complex constructs and require more computational steps. Many features also rely on adjustable parameters, such as the intensity discretization scheme, prior to texture calculations. Nonetheless, it is up to research teams to define the set of radiomic features and associated extraction parameters most relevant to address a given clinical question. With multiple options for image preprocessing steps and the flexibility of extraction parameters, a given radiomic feature set can easily reach a size close to $\sim$1,000 features only when considering the original image intensities \cite{radiomics_1k_5}. If filtering is considered prior to feature extraction, the feature set size could rise to $\sim$10,000 and more \cite{radiomics_10k_6}. Overall, the variability in feature types requiring increasing computational steps gives rise to the concept of complexity of radiomic feature extraction.

Furthermore, radiomics studies often integrate machine learning processes aiming at building predictive models from these extracted features with the goal, for example, to classify tumors and predict patient outcomes. The high dimensionality of radiomic feature sets introduces yet another layer of complexity to the overall radiomic analysis pipeline, necessitating careful feature selection and model optimization to avoid overfitting and ensure generalizability. Consequently, more emphasis should be given to focusing on the most effective features for a given clinical question. For instance, in an overall survival prediction in lung cancer \cite{vulnerabilities_8}, it was found that the tumor volume alone was driving the prognostic accuracy and that intensity and texture values were not relevant for prognostication. On the other hand, more complex features, such as gray-level histograms combined with texture features extracted from high-resolution CT images, following wavelet transforms, demonstrated a high accuracy in identifying lung tissue types \cite{depeursinge_lung_9}. This suggests that certain types of features are more suitable than others for predictive modeling in specific radiological clinical questions. Consequently, there is a need for methodologies capable of identifying the smallest set of predictive features while maintaining strong predictive performance and minimizing analytical complexity.

\textit{MEDiml} was developed to address this challenge. Although the platform supports the extraction of thousands of radiomic features, its analytical framework is specifically designed to identify compact and informative feature subsets. By systematically removing non-informative and redundant features, \textit{MEDiml} focuses on retaining only those that contribute meaningfully to a given clinical task, thereby reducing model complexity while improving interpretability and maintaining predictive performance (Figure \ref{fig:introductory}, Block 2). The remainder of this section is structured as follows: Subsection \ref{sec:rad_feat_ext} describes radiomic feature extraction, Subsection \ref{sec:modeling} describes modeling, and Subsection \ref{sec:res_analysis_interp} describes results analysis and interpretation.

\begin{figure}
  \centering
  \includegraphics[scale=0.28]{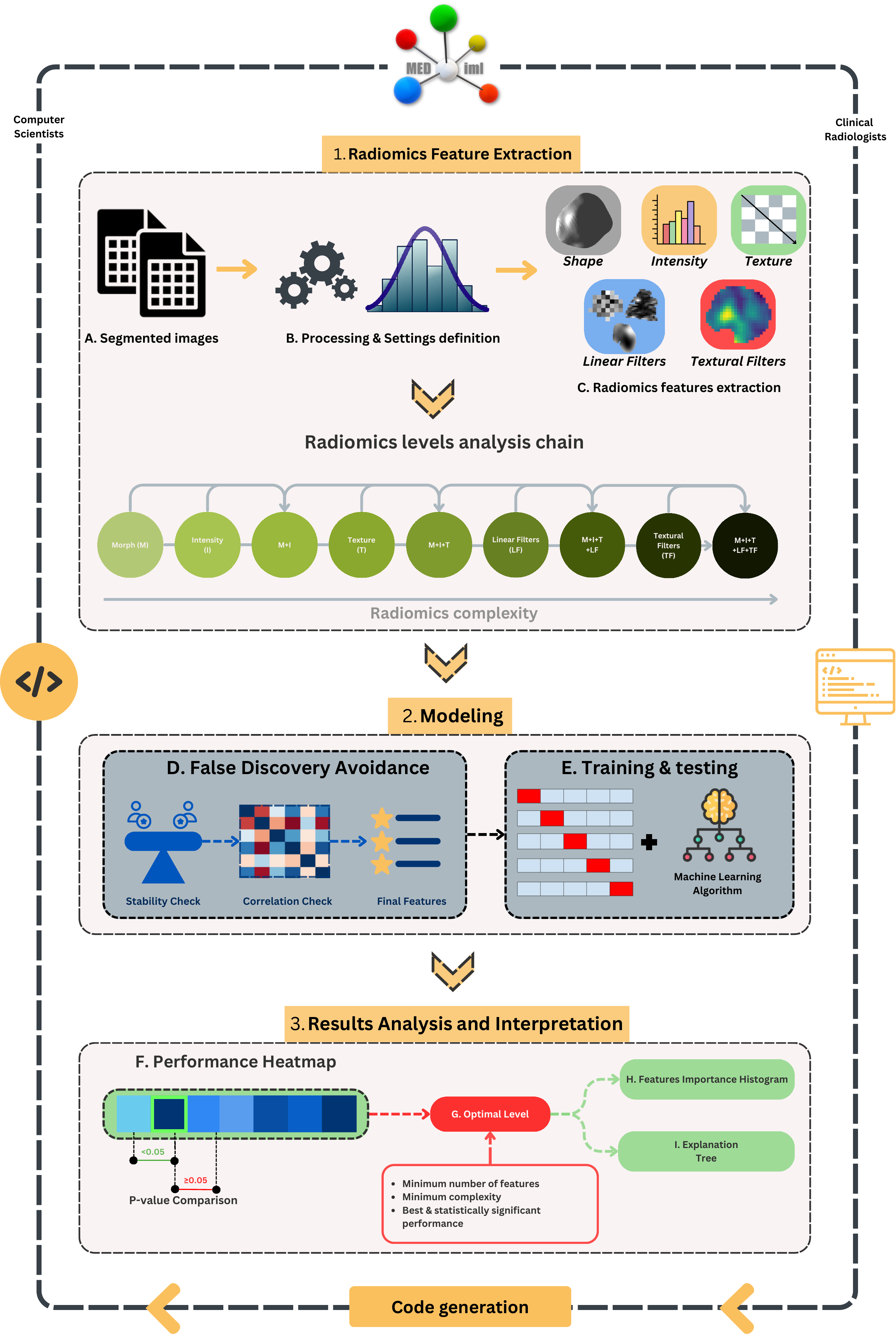}
  \caption{\textbf{Overview of the \textit{MEDiml} workflow.} The workflow in a typical radiomics analysis starts with acquisition and reconstruction of medical images. Subsequently, images are segmented to define regions of interest (ROIs). Following this step, the proposed radiomics software processes the images and computes features characterizing the ROIs, which are then organized by complexity levels for model training. Machine learning begins with feature cleaning to remove or replace invariant features, followed by feature set reduction to retain features exhibiting a high and stable correlation with the clinical endpoint, while removing inter-correlated features. All models are constructed using XGBoost \cite{chen_xgboost_2016}. The final step involves results analysis through two stages: identification of the optimal complexity level, characterized by the minimum number of features, minimum complexity, and the highest and statistically significant performance, and explanation based on feature importance. Experiments can be conducted via programming or through the \textit{MEDiml} interface, with the code generation option facilitating the shift between the two approaches.}
  \label{fig:workflow}
\end{figure}

\subsection{Radiomics Feature Extraction}
\label{sec:rad_feat_ext}
Radiomic feature extraction constitutes the first stage of the \textit{MEDiml} workflow (Figure \ref{fig:workflow}.1). This process begins with the preprocessing of raw medical images, followed by the specification of extraction parameters tailored to each feature category. The features are then extracted through a parallelized batch-processing framework, using \textit{MEDiml}’s \textit{BatchExtractor} class in the package or the \textit{BatchExtractor} page in the application, generating an individual JavaScript Object Notation (JSON) file for each scan. Finally, all extracted features are grouped into a single comma-separated values (CSV) file representing the entire dataset, which can be directly used in the machine learning step.

\subsection{Modeling}
\label{sec:modeling}
Predictive modeling constitutes the second stage of the \textit{MEDiml} workflow (Figure \ref{fig:workflow}.2). In radiomics, predictive modeling refers to the use of quantitative imaging features, together with statistical and machine learning methods, to predict clinical outcomes, characterize disease phenotypes, or support diagnostic and prognostic decision-making. Before model development, the extracted radiomic features undergo a data curation process to ensure the quality and reliability of the input data. This process includes the removal of problematic or non-informative features, such as invariant features that hold identical values across all samples, features with excessive missing values, and highly correlated features that may introduce redundancy. Feature normalization is subsequently applied to standardize the scale of the variables, ensuring that differences in feature magnitude do not bias the learning process and that all features contribute comparably to model training.

\subsubsection{Feature Set Reduction (FSR)}
\label{sec:fsr}
As discussed in Section \ref{sec:mediml_workflow}, radiomics studies can generate extremely high-dimensional datasets, often consisting of several thousand features and, in some cases, more than 10,000. While such large feature sets provide a rich quantitative description of medical images, they also increase the risk of overfitting, false discoveries, and reduced model generalizability. Consequently, FSR, also known as feature selection, is a critical step in the predictive modeling workflow (Figure \ref{fig:workflow}.2.D). The objective of FSR is to identify the subset of features that contains the most relevant information for a given predictive task while eliminating redundant, irrelevant, or noisy variables. By reducing the dimensionality of the dataset, feature selection improves model interpretability, decreases computational complexity, and often enhances predictive performance.

FSR relies on statistical and machine learning techniques to evaluate the relationship between features and the outcome of interest. In addition, FSR aims to remove highly correlated features that convey overlapping information, thereby reducing redundancy within the feature space. Feature selection methods are generally categorized into three families: filter, wrapper, and embedded approaches \cite{jovic_review_2015}.

Filter methods perform feature selection as a preprocessing step prior to model training and are independent of the learning algorithm used. These methods evaluate features based on statistical criteria, such as correlation coefficients, mutual information, or hypothesis-testing measures. For example, Spearman correlation can be used to identify groups of highly correlated features, allowing redundant variables to be removed while preserving representative features from each group. Filter methods are computationally efficient and scalable to large radiomics datasets; however, because they do not consider the downstream learning algorithm, they may not always identify the optimal subset of features for a specific predictive task.

Wrapper methods, in contrast, evaluate feature subsets directly through model performance. These approaches employ search strategies to explore different combinations of features and identify the subset that yields the best predictive results according to a predefined evaluation metric. Common wrapper-based techniques include sequential forward selection, which progressively adds features to the model, and sequential backward elimination, which iteratively removes features from an initially complete feature set. While wrapper methods can produce highly effective feature subsets, they are computationally expensive, particularly when applied to the large feature spaces commonly encountered in radiomics.

Embedded methods integrate feature selection directly into the model training process, combining many of the advantages of both filter and wrapper approaches. In these methods, feature importance is learned as part of the optimization procedure of the machine learning algorithm itself. Examples include Least Absolute Shrinkage and Selection Operator (LASSO) regression \cite{tibshirani_lasso_1996}, which performs feature selection through regularization, as well as tree-based methods such as Random Forest \cite{breiman_rf_2001} and eXtreme Gradient Boosting (XGBoost) \cite{chen_xgboost_2016}, which estimate feature importance during model construction. Embedded methods generally provide a favorable balance between computational efficiency, predictive performance, and resistance to overfitting, making them popular choices in radiomics applications.

In \textit{MEDiml}, FSR is implemented through the \textit{FSR} class and within the \textit{Feature Reduction} node. The current implementation employs a filter-based strategy that uses correlation statistics to identify and remove redundant features, retaining those that are most informative for the predictive modeling task under investigation. This approach provides an efficient and scalable solution for reducing the dimensionality of large radiomics datasets while preserving their most relevant predictive information.

\subsubsection{Training and Testing}
\label{sec:train_and_test}
Before fitting data-dependent preprocessing, the dataset is typically divided into training and testing subsets to enable an unbiased evaluation of model performance. In most cases, techniques such as cross-validation \cite{arlot_cv_2010} or random subsampling are employed to improve the robustness of model selection and reduce the risk of overfitting. A machine learning algorithm, such as Random Forest \cite{breiman_rf_2001}, Support Vector Machine \cite{cortes_svm_1995}, or Logistic Regression \cite{lavalley_logreg_2008}, is then trained on the curated radiomic feature set to learn patterns associated with the target clinical outcome (Figure \ref{fig:workflow}.2.E). During training, the model identifies relationships between imaging-derived features and known labels or outcomes, allowing it to make predictions on previously unseen data. In each training iteration, models are retrained multiple times to select the best-performing model parameters. A key difference in \textit{MEDiml}’s modeling approach is training a model for each radiomic complexity level, from the least complex level to the most complex, i.e., comparing models trained on progressively more complex radiomic feature sets in order to identify the simplest feature set that achieves the highest predictive performance \cite{loutfi_complexity_2024}. A statistical significance test is then used to determine the optimal level, defined as the level yielding the highest predictive performance, with the least complexity, for the given clinical problem. This enables researchers to focus on simpler models and avoid unnecessary complexity. This can be done in \textit{MEDiml} through the \textit{RadiomicsLearner} class or the \textit{Radiomics Learner} node.

\subsection{Results Analysis and Interpretation}
\label{sec:res_analysis_interp}
Results analysis and interpretation constitute the final stage of the \textit{MEDiml} workflow (Figure \ref{fig:workflow}.3). The predictive performance of the resulting model is assessed using appropriate evaluation metrics, such as accuracy, sensitivity, specificity, and the Area Under the receiver operating characteristic Curve (AUC) \cite{fawcett_roc_2006}. Through this process, predictive modeling transforms quantitative radiomic data into clinically meaningful insights that can support precision medicine and evidence-based decision-making. However, predictive performance alone is insufficient for model selection, as observed differences between models may arise from sampling variability rather than genuine performance differences. Consequently, statistical significance testing is required to determine whether one model meaningfully outperforms another. Depending on the evaluation setting, commonly used approaches include paired t-test \cite{student_ttest_1908}, \textit{Wilcoxon signed-rank test} \cite{wilcoxon_1945}, \textit{Nadeau-Bengio corrected t-test} \cite{nadeau_bengio_2003}, and \textit{DeLong's test} \cite{delong_1988, delong_24} for comparing AUC values. Within \textit{MEDiml}, predictive models generated from different radiomic complexity levels are systematically compared using these statistical methods to identify the optimal balance between predictive performance and model complexity (Figure \ref{fig:workflow}.3.F). In particular, \textit{DeLong's test} is used for AUC-based comparisons, while other tests such as \textit{paired t-test} and \textit{Wilcoxon signed-rank test} are available. This framework, handled by the \textit{Results} and \textit{Stats} classes and the \textit{Analyze} node, enables the selection of the lowest-complexity radiomic model that achieves statistically equivalent or superior predictive performance without a detected statistically significant gain at higher complexity (Figure \ref{fig:workflow}.3.G), thereby simplifying predictive modeling and promoting model interpretability.

\section{The \textit{MEDiml} package}
\label{sec:mediml_pkg}
Recall that the \textit{MEDiml} Python package (\textit{MEDiml-pkg}) provides the computational framework used by both programmatic analyses and the graphical application. The workflow described in Section \ref{sec:mediml_workflow} is implemented through its modules and classes, whose roles, features and configuration options are detailed below (Figure \ref{fig:introductory}, Block 3, panel C).

\subsection{Overview}
\label{sec:mediml_pkg_overview}
The package is compliant with IBSI phases 1 and 2 \cite{ibsi_1, ibsi_2} and supports parallelized batch feature extraction. The implementation builds upon widely adopted scientific Python libraries, including \textit{NumPy} \cite{harris_numpy_2020} for numerical computation, \textit{Pandas} \cite{pandas_2020} for data manipulation, and \textit{PyCaret} \cite{ali_pycaret_2024} for machine learning workflows, among others. In contrast to conventional radiomics frameworks that prioritize the extraction and the analysis of the full feature sets, \textit{MEDiml} is designed around the radiomics complexity concept introduced earlier. It implements a complexity-minimization approach that emphasizes the identification of compact, informative feature subsets capable of achieving strong predictive performance while maintaining model interpretability.

The package guides users through a standardized end-to-end workflow, from medical images to result interpretation (Figure \ref{fig:workflow}). The \textit{DataManager} class loads images and their segmentation masks (Subsection \ref{sec:data_wrangling}), after which the processing subpackage prepares the images for feature extraction (Subsection \ref{sec:img_processing}). Optional filtering is provided by the filters subpackage (Subsection \ref{sec:filtering}), and the \textit{BatchExtractor} class in the biomarkers subpackage extracts radiomic features (Subsection \ref{sec:biom_ext}). The \textit{FSR} class reduces the feature sets (Subsection \ref{sec:pkg_fsr}), and \textit{RadiomicsLearner} trains and evaluates models across radiomic complexity levels (Subsection \ref{sec:pkg_modeling}). Finally, the \textit{Results} class supports model comparison and interpretation (Subsection \ref{sec:model_interp}), followed by model finalization using \textit{RadiomicsLearner} (Subsection \ref{sec:model_finalization}).

\subsection{Data Wrangling}
\label{sec:data_wrangling}
The radiomics workflow begins with data loading through the \textit{DataManager} class in the wrangling subpackage. \textit{MEDiml} currently supports the two most common medical imaging formats used in research and clinical environments: Digital Imaging and Communications in Medicine (DICOM) and Neuroimaging Informatics Technology Initiative (NIfTI). Support for additional formats, including Nearly Raw Raster Data (NRRD), is planned for future releases.

During data loading, imaging scans are automatically associated with their corresponding segmentation masks, ensuring that regions of interest (ROIs) are correctly linked to image volumes. Relevant image metadata are simultaneously extracted and made available throughout the analysis pipeline. To facilitate quality control and parameter selection, the module provides exploratory analyses of the dataset, including voxel intensity distribution statistics and histogram visualizations that assist users in selecting appropriate preprocessing and feature extraction settings.

\subsection{Image Processing}
\label{sec:img_processing}
Following data loading, images are processed using methods compliant with the recommendations of the IBSI \cite{ibsi_1}, implemented in the processing subpackage. Because radiomic features can be highly sensitive to image acquisition and reconstruction parameters, standardized preprocessing is critical for ensuring reproducibility and comparability across studies.
\textit{MEDiml} implements all major IBSI-defined preprocessing operations, including:
\begin{itemize}
\item Interpolation and resampling;
\item ROI extraction;
\item Re-segmentation;
\item Intensity discretization;
\item Standardized Uptake Value (SUV) computation for Positron Emission Tomography (PET) imaging.
\end{itemize}

To accommodate different study designs, each processing step offers several options the user can select from. For example, in interpolation, multiple algorithms are available, including nearest-neighbor, linear, spline, and cubic interpolation. These preprocessing operations allow users to harmonize imaging data prior to feature extraction.

\subsection{Filtering}
\label{sec:filtering}
Image filtering constitutes an optional stage of the workflow that can be applied prior to feature extraction. The objective of filtering is to enhance specific image characteristics, reduce noise, and generate derived image representations that may contain additional predictive information not readily apparent in the original image.

\textit{MEDiml} currently supports five classes of image filters, using the filters subpackage:

\begin{itemize}
\item Mean;
\item Laplacian-of-Gaussian (LoG);
\item Laws;
\item Gabor;
\item Wavelet.
\end{itemize}

\subsubsection{Nonlinear Filtering}
\label{sec:non_linear_filtering}
\textit{MEDiml}’s filters subpackage also implements, through its \textit{TexturalFilter} class, nonlinear filtering accelerated by a graphics processing unit (GPU) \cite{buck_cuda_2007, klockner_pycuda_2012}, referred to here as textural filtering. In this approach, a predefined three-dimensional cubic kernel is iteratively moved throughout the image volume, and texture features are calculated locally at each spatial position. Currently, only filtering using the gray-level co-occurrence matrix (GLCM) \cite{sebastian_glcm_2012} is supported. It generates 25 textural filter maps corresponding to the 25 texture features defined within the GLCM family. These derived images can subsequently be used as inputs for radiomic feature extraction, enabling the characterization of texture patterns at multiple spatial scales.

\subsection{Biomarker Extraction}
\label{sec:biom_ext}
Radiomic biomarker extraction represents the central component of the \textit{MEDiml} workflow. Extraction is performed through the \textit{BatchExtractor} class within the biomarkers subpackage, which supports both single-patient and cohort-level analyses. To accelerate computation on large imaging datasets, extraction tasks are parallelized using the \textit{Ray} framework \cite{moritz_ray_2018}.
\textit{MEDiml} supports the extraction of all major IBSI-compliant \cite{ibsi_1} radiomic feature families:
\begin{itemize}
    \item Morphological features;
    \item Local intensity features;
    \item Intensity statistics features;
    \item Intensity histogram features;
    \item Intensity-volume histogram features;
    \item Gray-Level Co-occurrence Matrix (GLCM) features \cite{sebastian_glcm_2012};
    \item Gray-Level Run Length Matrix (GLRLM) features;
    \item Gray-Level Size Zone Matrix (GLSZM) features;
    \item Gray-Level Distance Zone Matrix (GLDZM) features;
    \item Neighborhood Gray Tone Difference Matrix (NGTDM) features;
    \item Neighboring Gray Level Dependence Matrix (NGLDM) features.
\end{itemize}

Feature extraction can be performed directly on original images or on filtered image representations, resulting in a high-dimensional set of biomarkers spanning multiple levels of radiomic complexity.

\subsection{Feature Set Reduction: The False Discovery Avoidance (FDA) method}
\label{sec:pkg_fsr}
Because radiomic studies routinely generate thousands of candidate features while operating on relatively small patient cohorts, feature reduction is an essential step of the analysis workflow. \textit{MEDiml} addresses this challenge using an implementation of the FDA method \cite{fda_21} within the \textit{Feature Set Reduction (FSR)} class of its learning subpackage.
At each radiomic complexity level and for every training fold, candidate features are ranked according to their correlation with the clinical outcome while simultaneously controlling feature redundancy. Stability is estimated through repeated stratified random subsampling, using a configurable number of training subsamples (100 by default) and calculating Spearman correlation \cite{peng_mrmr_2005} coefficients across all repetitions. For each feature, both the minimum correlation coefficient ($R_{\mathrm{min}}$) and the mean correlation coefficient ($R_{\mathrm{mean}}$) are computed.

\begin{figure}
  \centering
  \includegraphics[scale=0.172]{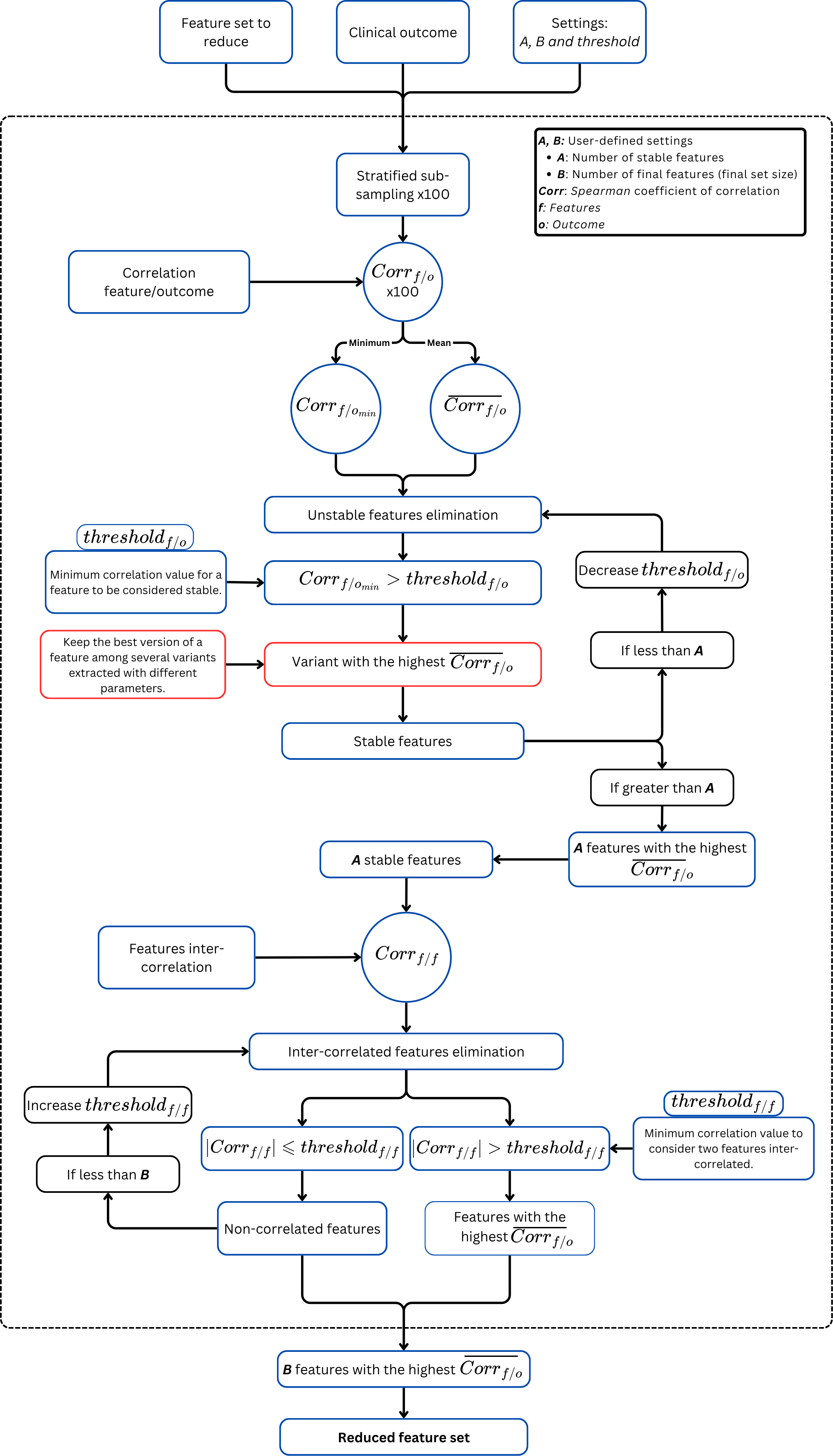}
   \caption{\textbf{Overview of the False Discovery Avoidance method.} The procedure ranks candidate features by their \textit{Spearman correlation} with the clinical outcome and removes unstable or redundant features before model training.}
  \label{fig:fda}
\end{figure}

Features with an absolute $R_{\mathrm{min}}$ greater than 0.5 are considered stable and retained for further analysis. When multiple variants of the same feature are available, such as those generated using different extraction parameters, only the variant exhibiting the highest $R_{\mathrm{mean}}$ is retained. To further reduce redundancy, pairs of features exhibiting an absolute intra-correlation greater than 0.7 are examined, and the feature displaying the weaker association with the clinical outcome is removed. This process is illustrated fully in Figure \ref{fig:fda}.

This procedure yields compact feature subsets, typically containing between 5 and 20 biomarkers, that exhibit strong outcome association while preserving diversity and stability.

\subsection{Modeling}
\label{sec:pkg_modeling}
The reduced feature subsets are subsequently passed to the learning subpackage for predictive modeling. This subpackage supports the complete machine learning workflow, including dataset partitioning, feature cleaning, normalization, feature reduction, model training, and model evaluation.
The framework provides access to the collection of machine learning algorithms available through \textit{PyCaret} \cite{ali_pycaret_2024}, allowing users to evaluate diverse modeling approaches to select the best one for their workflow. Model validation can be performed using several commonly adopted evaluation schemes, including:
\begin{itemize}
    \item $k$-fold cross-validation \cite{arlot_cv_2010}.
    \item random subsampling.
\end{itemize}
The main goal of the learning subpackage is to enable the systematic assessment of whether increasing feature count improves predictive performance.

\subsection{Results Analysis and Interpretation}
\label{sec:model_interp}
The final stage of the \textit{MEDiml} workflow focuses on model comparison, statistical validation, and interpretability, also managed by the learning subpackage, using the \textit{Results} class. Performance differences observed between competing models may arise from random variation rather than genuine predictive superiority. Consequently, \textit{MEDiml} incorporates multiple statistical significance testing procedures to compare models trained using different radiomic complexity levels. Supported methods include:
\begin{itemize}
    \item \textit{DeLong's test} \cite{delong_1988, delong_24} for AUC \cite{fawcett_roc_2006} comparison.
    \item \textit{Wilcoxon signed-rank tests} \cite{wilcoxon_1945}.
    \item \textit{Paired t-test}s \cite{student_ttest_1908}.
    \item \textit{Nadeau-Bengio corrected t-test} \cite{nadeau_bengio_2003}.
\end{itemize}

These tests allow the identification of the optimal complexity level by determining whether increases in complexity yield statistically significant improvements in predictive performance.

\begin{figure}
  \centering
  \includegraphics[scale=0.572]{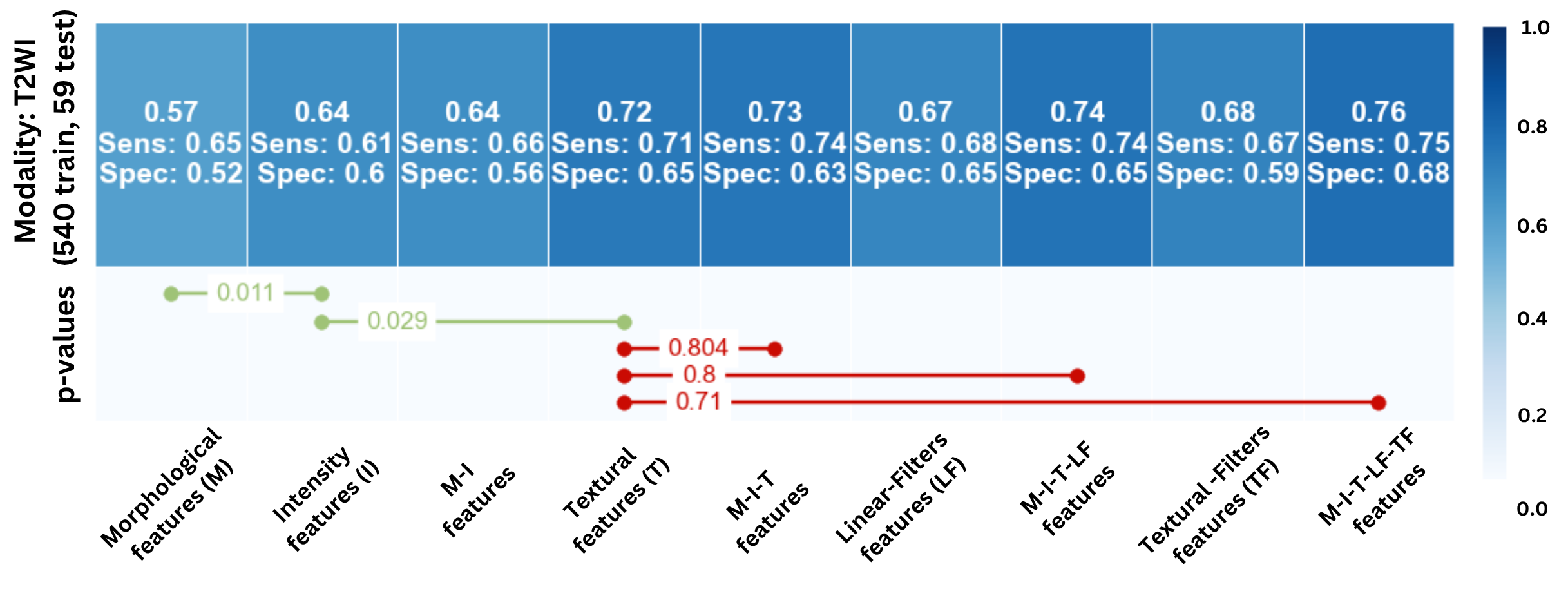}
  \caption{\textbf{Mean AUC performance heatmap across testing fold for each radiomic complexity level.} Each column indicates a complexity level. The numerical suffixes in column names indicate the number of features retained after feature set reduction. In the bottom rows, green lines indicate significant improvements (p < 0.05) and red lines indicate non-significant improvements. Sens: Sensitivity; Spec: Specificity.}
  \label{fig:heatmap}
\end{figure}

To facilitate model selection, \textit{MEDiml} generates performance heatmaps that summarize predictive metrics and statistical significance across complexity levels (Figure \ref{fig:heatmap}). The optimal level is defined as the complexity level that achieves strong predictive performance while minimizing feature count and complexity.

\begin{figure}
  \centering
  \includegraphics[scale=0.272]{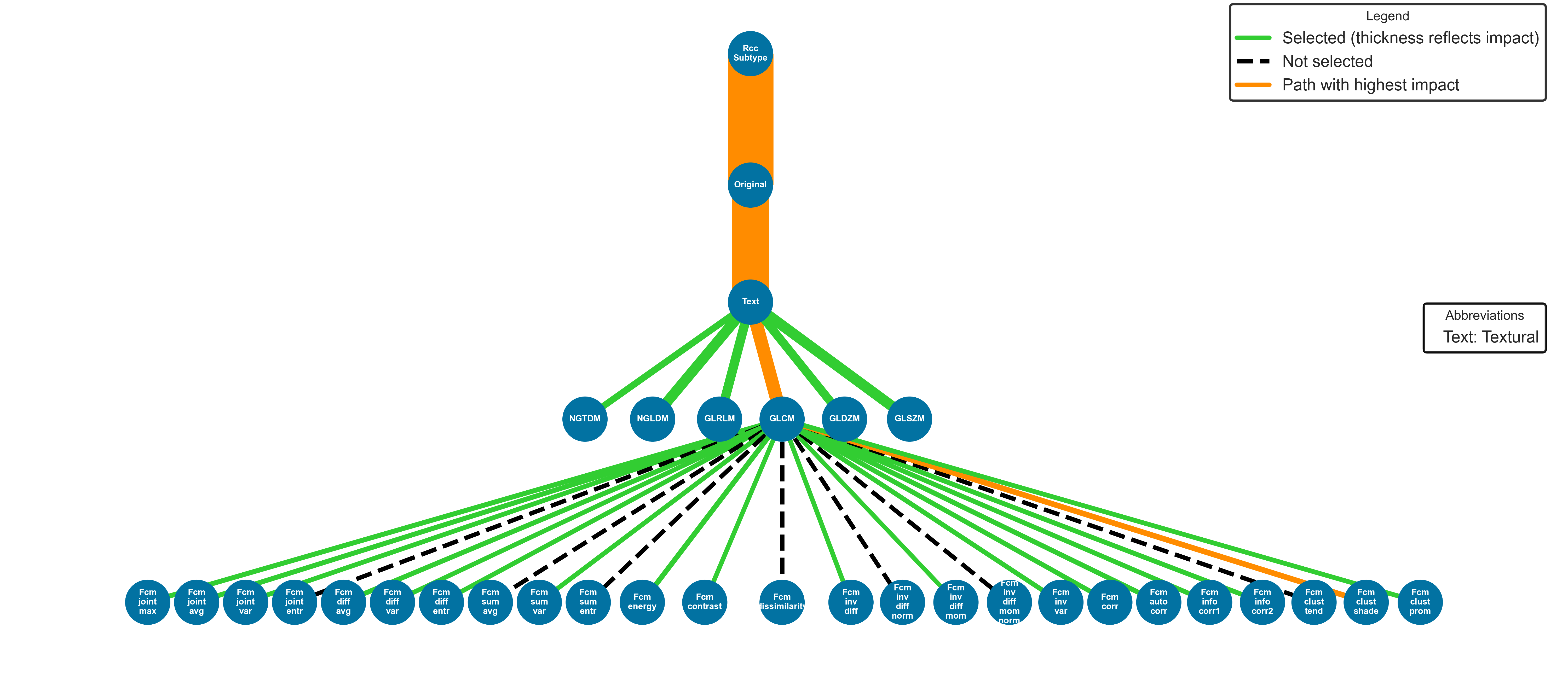}
  \caption{\textbf{Explanation Tree for textural-filter features in renal cell carcinoma subtype classification.} The orange line draws the path leading to the most predictive feature in the set. Green lines represent other features that participated in the model’s decision-making, with line thickness reflecting their contribution to the predictive performance (i.e., the thicker the line, the higher the feature importance). In this example, the orange line points to the GLCM cluster shade feature, indicating that it contributed most to the model's decision-making.}
  \label{fig:explanation_tree}
\end{figure}

\begin{figure}
  \centering
  \includegraphics[scale=0.172]{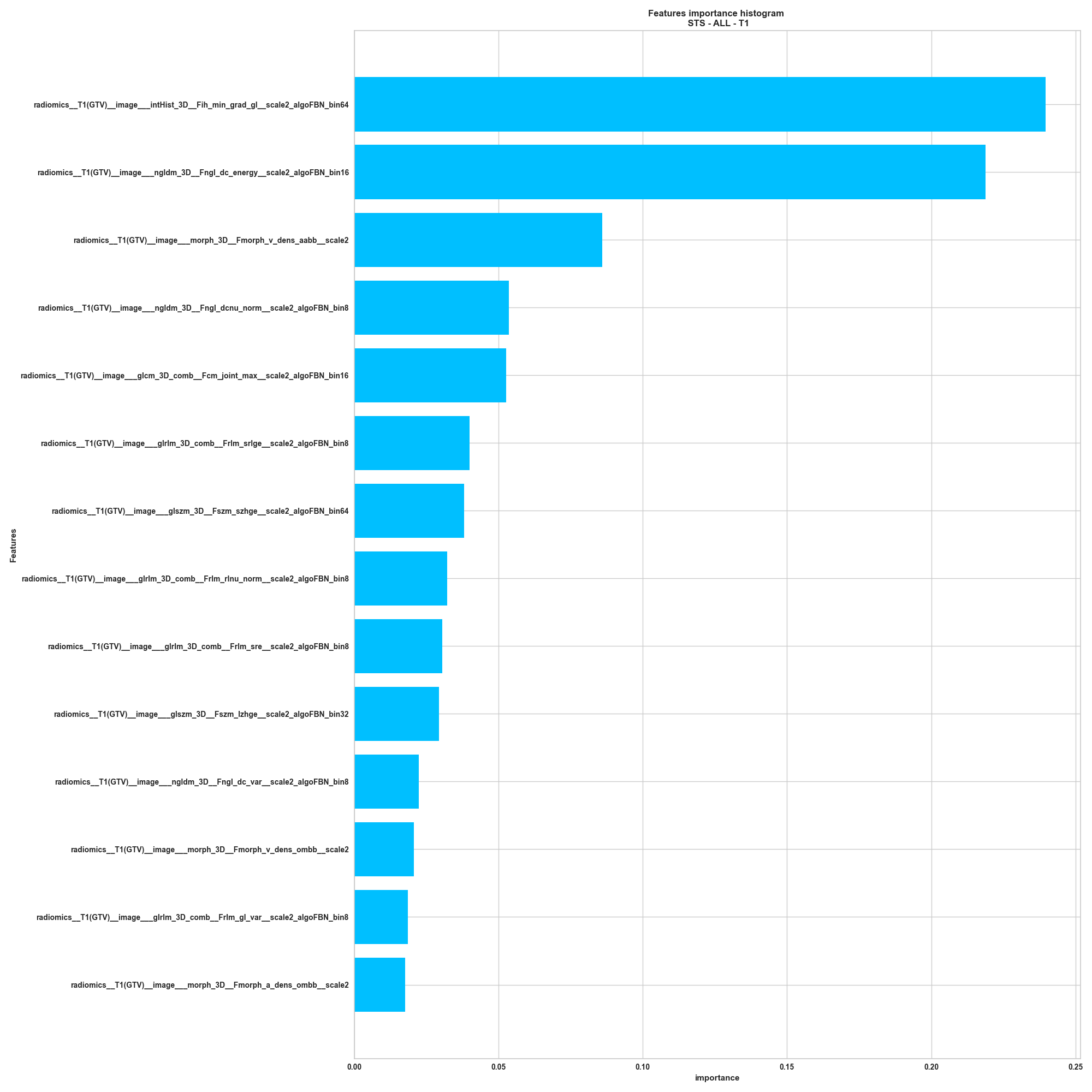}
   \caption{\textbf{Feature-importance histogram for clear cell versus non-clear cell renal cell carcinoma classification.} Features are sorted by their importance, from highest to lowest, i.e., the longer the blue bar, the higher the importance.}
  \label{fig:histogram}
\end{figure}

For model interpretation, \textit{MEDiml} relies on model-specific feature importance scores but, most importantly, provides a visualization method called the Explanation Tree (Figure \ref{fig:workflow}.3.I and Figure \ref{fig:explanation_tree}). The hierarchical grouping of model-specific importance scores is an additional contribution to \textit{MEDiml}’s model-interpretation tools. This hierarchical representation decomposes model decisions by tracing predictive contributions from the selected complexity level through successive layers of abstraction, including image filters when applicable, radiomic feature families, and individual features. Branch widths are proportional to feature importance, and the path leading to the most influential biomarker is emphasized visually. Complementary feature-importance histograms (Figure \ref{fig:workflow}.3.H and Figure \ref{fig:histogram}) are also provided to quantify the relative contribution of each selected feature.

\subsection{Model Finalization}
\label{sec:model_finalization}
The final stage of the \textit{MEDiml} workflow consists of model finalization, where the results of the complexity-level comparison and statistical analyses are used to identify the optimal predictive pipeline. Once identified, the optimal pipeline is retrained using the entire learning dataset to maximize the amount of information available for model construction. The resulting trained model, together with all preprocessing, feature extraction, feature reduction, and learning parameters, is then serialized and saved for future deployment. If a holdout set is provided, a final independent evaluation is performed and the corresponding predictive metrics are reported. This final validation can provide an unbiased estimate of the model's performance on previously unseen data \cite{cawley_overfitting_2010} and serves as the last verification step before deployment.

Model finalization therefore transforms the exploratory analyses conducted throughout the \textit{MEDiml} workflow into a single validated predictive model that can be applied to new datasets while preserving the exact methodological choices that led to its selection. This stage completes the end-to-end \textit{MEDiml} radiomics pipeline, from image acquisition and feature extraction to model interpretation, validation, and deployment-ready model generation.

Together, these analysis tools enable researchers to move beyond predictive performance alone and gain a deeper understanding of the radiomic biomarkers driving model behavior and decision-making, thereby improving interpretability and supporting the clinical translation of radiomics-based predictive models.

\section{The \textit{MEDiml} application}
\label{sec:mediml_app}
Recall that the \textit{MEDiml} application (\textit{MEDiml-app}) is the desktop graphical user interface (GUI) through which researchers can use the radiomics workflow without programming expertise. It uses the Python package (\textit{MEDiml-pkg}) described in Section \ref{sec:mediml_pkg}, with analyses assembled visually from connected operations. Understanding the application therefore requires familiarity with its workflow concepts and with the modules used to configure analyses (Figure \ref{fig:introductory}, Block 3, panel A).

\subsection{Overview}
\label{sec:medapp_overview}
\begin{figure}
  \centering
  \includegraphics[scale=0.272]{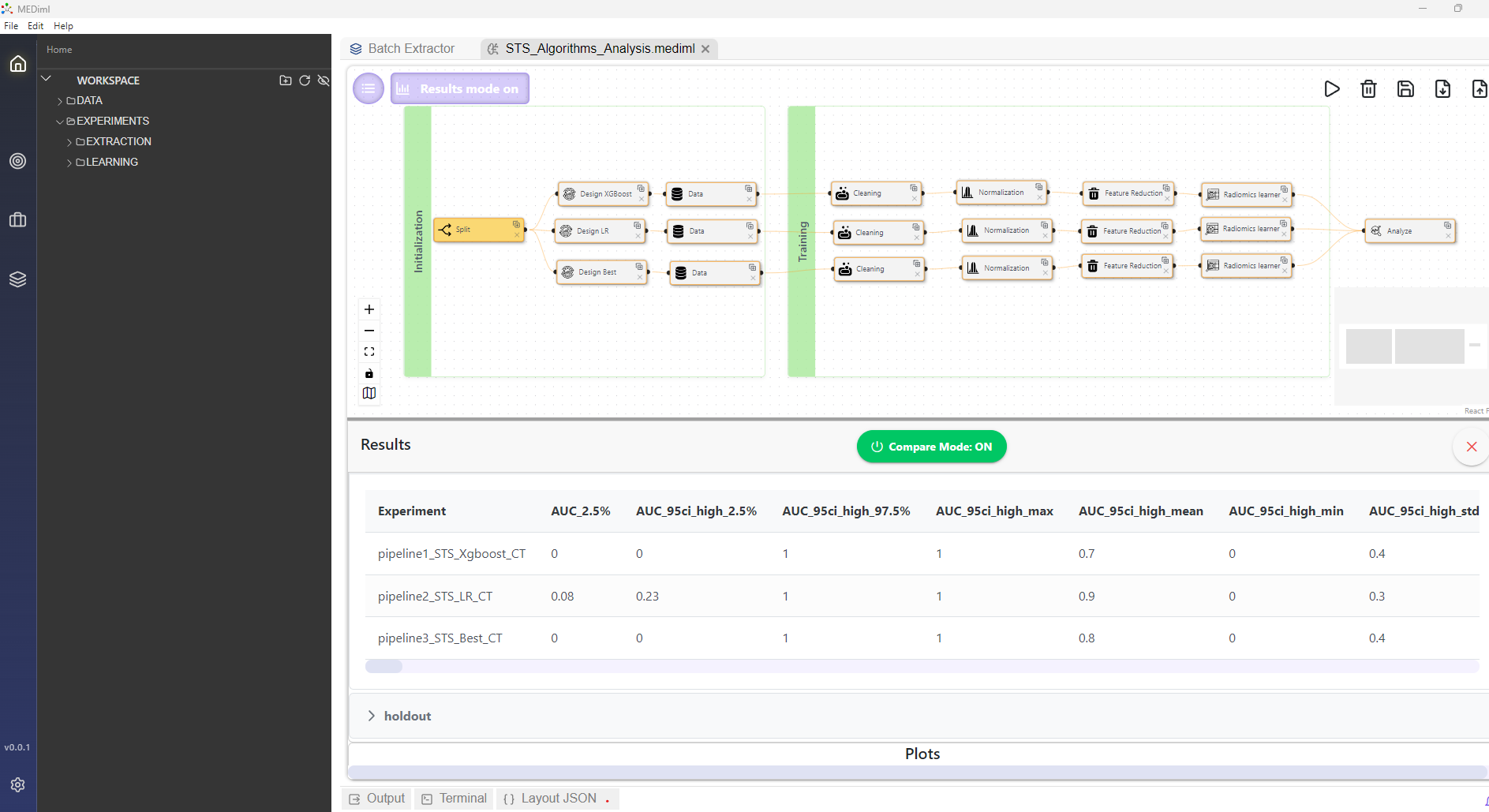}
  \caption{\textbf{The \textit{MEDiml} graphical interface,} showcasing the \textit{Learning} module, which provides a visual workspace for configuring and conducting machine learning analyses.}
  \label{fig:mediml_gui}
\end{figure}

Built around a drag-and-drop, node-based architecture, the application enables users to design, configure, and execute radiomics experiments entirely through visual interaction (Figure \ref{fig:mediml_gui}). At a computational level, every action available through the desktop application invokes the same underlying Python methods; the graphical interface and Python package act as complementary interfaces to a single computational package rather than as independent software products, which facilitates collaboration between clinical researchers and computational scientists by allowing visually designed experiments to be transferred directly into reproducible Python workflows without requiring manual reimplementation.

The application is organized into four modules corresponding to distinct stages of radiomics analysis: \textit{Single-Scan Extraction}, \textit{DataManager}, \textit{BatchExtractor}, and Learning. The remainder of this section is structured as follows: Subsection \ref{sec:medapp_nodes} defines nodes, pipelines and scenes, and Subsection \ref{sec:medapp_modules} describes the application modules.

\subsection{Nodes, Pipelines and Scenes}
\label{sec:medapp_nodes}
The \textit{MEDiml} application is built around three hierarchical concepts: nodes, pipelines, and scenes.

\begin{figure}[!htbp]
  \centering
  \includegraphics[scale=0.472]{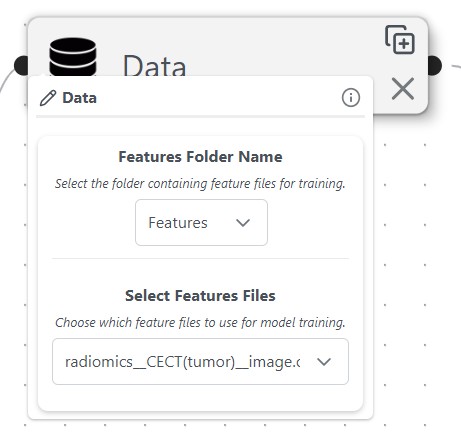}
  \caption{\textbf{Example of a node in the \textit{Learning} module.} The \textit{Data} node allows users to select the feature set(s) to use for their machine learning experiment.}
  \label{fig:data_node}
\end{figure}

A \textbf{node} represents the fundamental building block of the interface. Each node implements a specific computational operation, such as data loading, normalization, feature reduction, model training, or result analysis, and exposes a set of configurable parameters that can be adjusted prior to execution. Nodes contain well-defined inputs and outputs and can be connected to form multiple workflows. An example is presented in Figure \ref{fig:data_node}.

\begin{figure}[!htbp]
  \centering
  \includegraphics[scale=0.272]{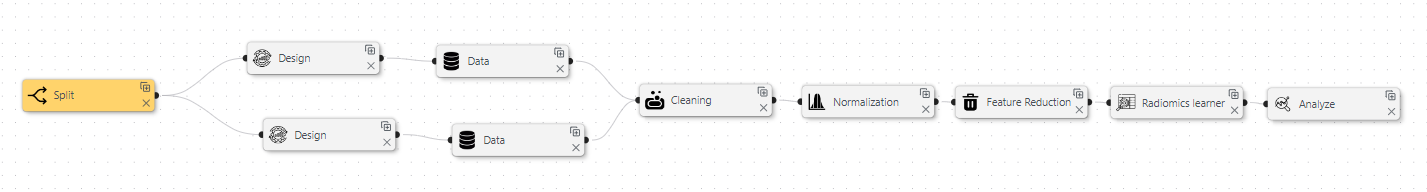}
   \caption{\textbf{Example of two machine learning pipelines.} Connected nodes define the operations used in an analysis for each pipeline.}
  \label{fig:ml_pipeline}
\end{figure}

A \textbf{pipeline} consists of an ordered sequence of connected nodes that collectively define a complete or partial radiomics workflow. Pipelines are assembled by selecting nodes from the available nodes list, placing them within the workspace, and defining their execution order through connections. Each pipeline represents a specific experimental configuration, such as a chosen radiomic complexity level, feature selection strategy, or machine learning model, and is always executed independently. An example is presented in Figure \ref{fig:ml_pipeline}.

\begin{figure}[!htbp]
  \centering
  \includegraphics[scale=0.182]{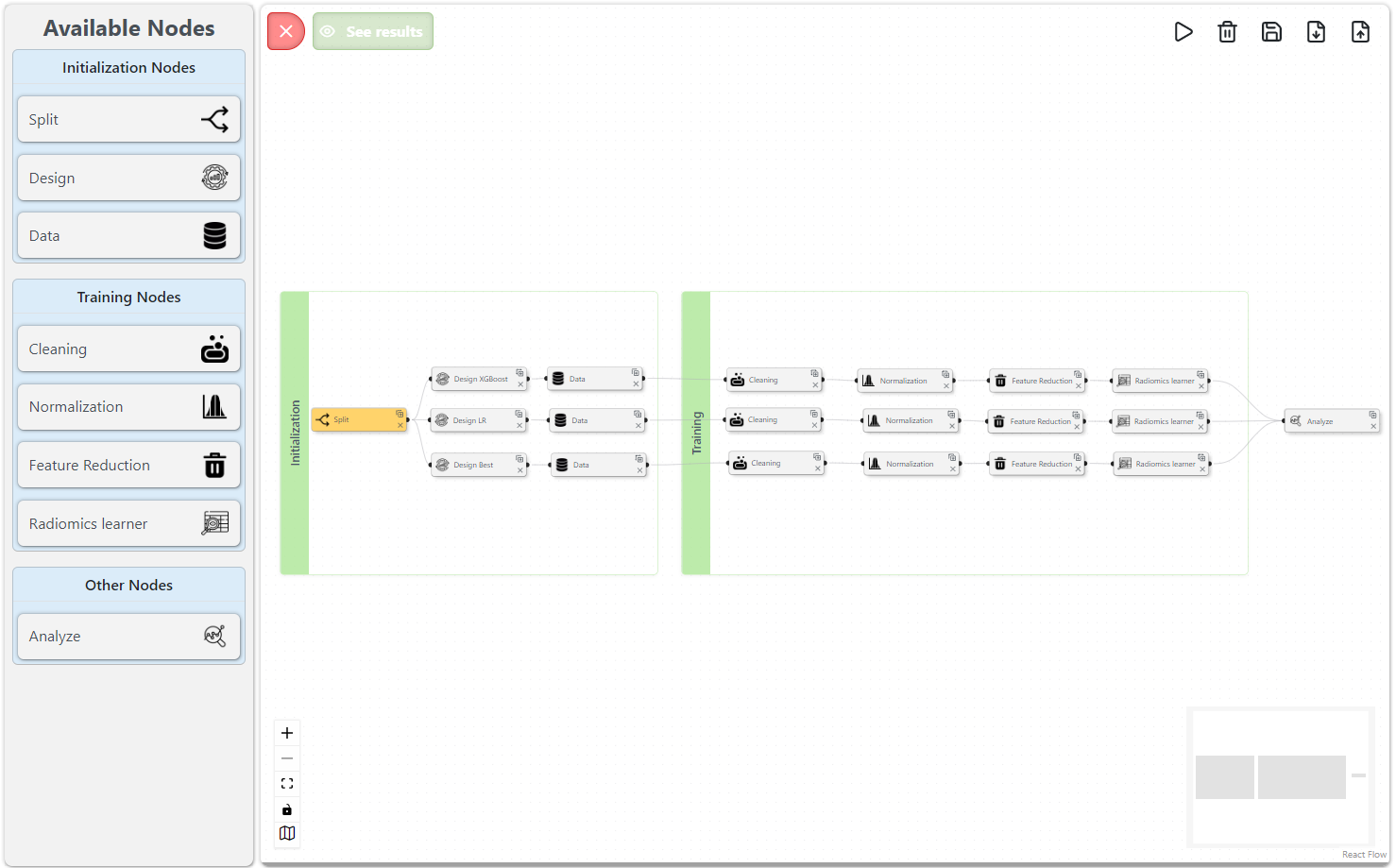}
   \caption{\textbf{Example of a \textit{Learning} module scene.} Multiple pipelines can be configured within the same scene to compare experimental settings.}
  \label{fig:ml_scene}
\end{figure}

A \textbf{scene} is the highest organizational level within the application and contains one or more pipelines. An example is presented in Figure \ref{fig:ml_scene}. Scenes enable users to design, execute, and compare multiple experiments simultaneously within a unified workspace. For example, separate pipelines may be constructed to evaluate different radiomic complexity levels on the same dataset. Scenes constitute the primary unit for saving, sharing, exporting, analyzing and reproducing analyses within \textit{MEDiml}.

\subsection{Modules}
\label{sec:medapp_modules}
The application modules follow the stages of a radiomics study, from the inspection of a single scan to model development on a full cohort. Subsection \ref{sec:medapp_single_scan} describes the \textit{Single-Scan Extraction} module, used to explore extraction settings on individual images. Subsection \ref{sec:medapp_dataManager} describes the \textit{DataManager} module, used to organize a dataset and guide the choice of extraction settings, and Subsection \ref{sec:medapp_batchExtractor} describes the \textit{BatchExtractor} module, used to extract radiomic features from an entire cohort. Finally, Subsection \ref{sec:medapp_learning} describes the \textit{Learning} module, used to train, compare and interpret predictive models from the extracted features.

\subsubsection{Single-Scan Extraction}
\label{sec:medapp_single_scan}
The \textit{Single-Scan Extraction} module is designed for exploratory radiomic feature extraction on individual medical images. Users can load a single DICOM or NIfTI image together with its corresponding segmentation mask, configure preprocessing and extraction parameters, and generate radiomic feature values for inspection.

\begin{figure}[!htbp]
  \centering
  \includegraphics[scale=0.24]{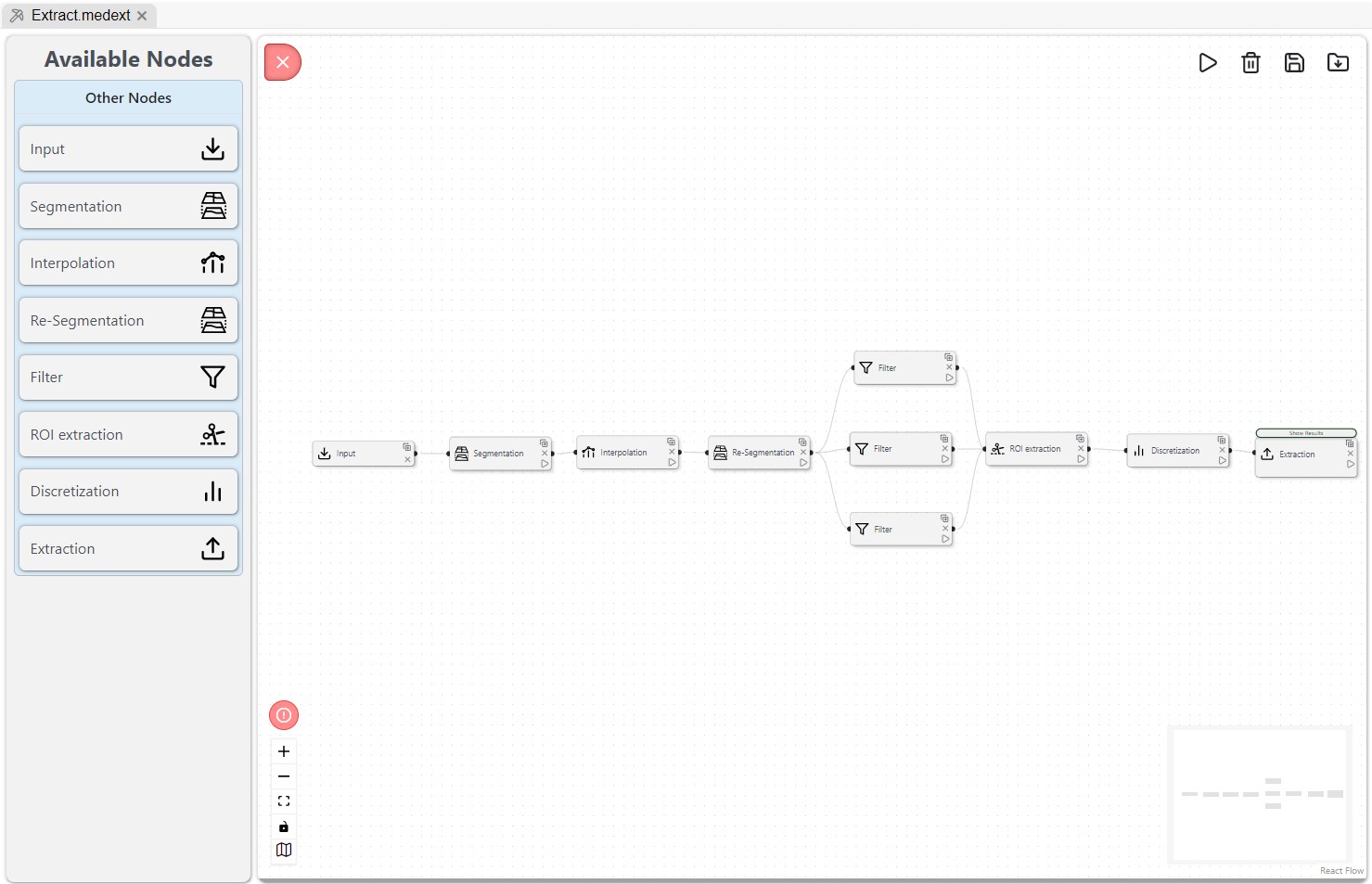}
   \caption{\textbf{Overview of the \textit{Single-Scan Extraction} module.} The scene contains three pipelines for comparing filter configurations on the same scan.}
  \label{fig:single_scan_scene}
\end{figure}

Available configuration options include interpolation settings, intensity re-segmentation ranges, and radiomic complexity levels. This module is particularly useful for validating extraction parameters, verifying feature extraction behavior on reference cases, testing different processing options, and familiarizing users with the effects of preprocessing choices before applying an analysis pipeline to an entire cohort. An overview of the \textit{Single-Scan Extraction} module is presented in Figure \ref{fig:single_scan_scene}.

\subsubsection{DataManager}
\label{sec:medapp_dataManager}
\begin{figure}[!htbp]
  \centering
  \includegraphics[scale=0.170]{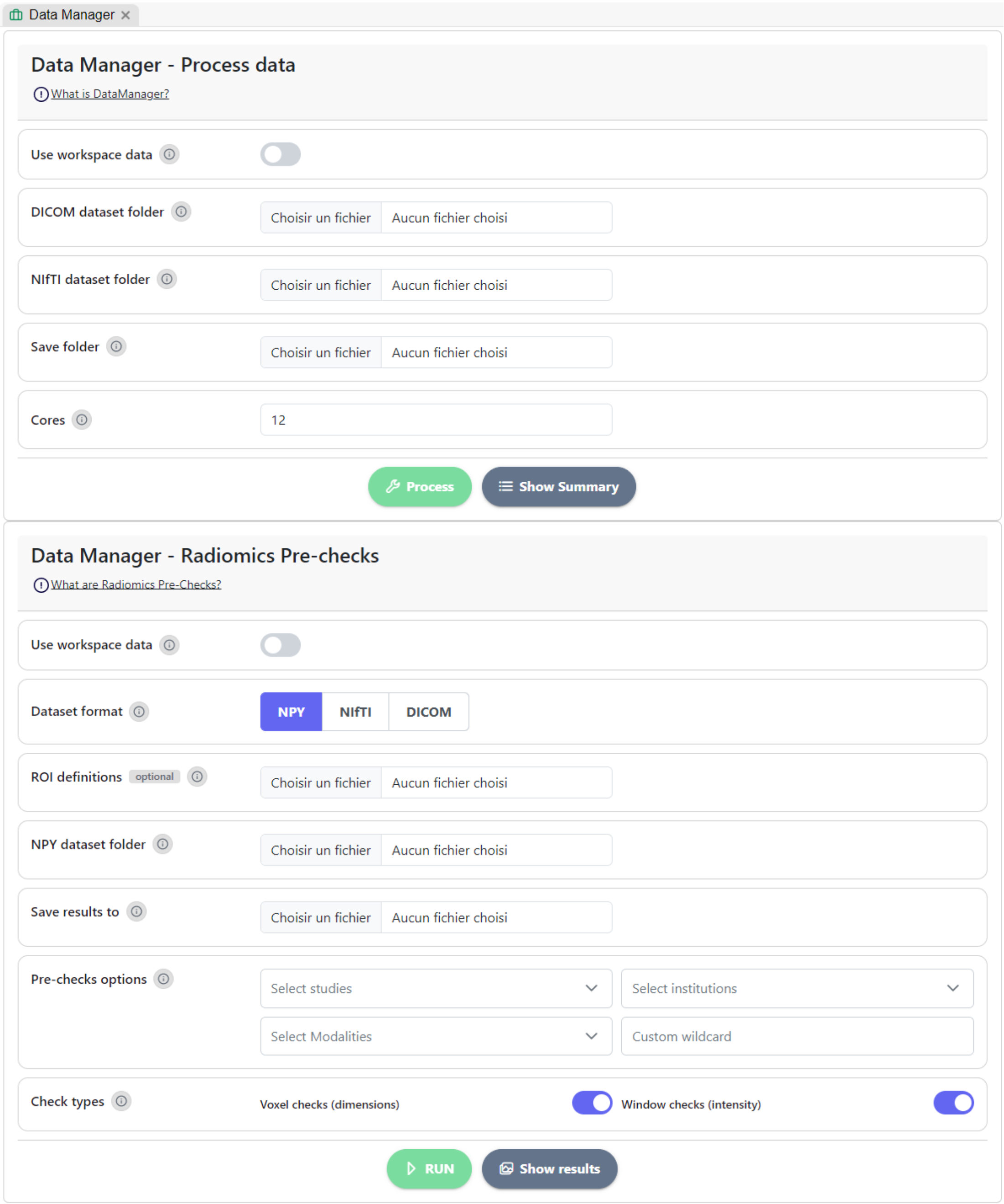}
  \caption{\textbf{Overview of the \textit{DataManager} module.} The \textit{Process data} step loads, reads and organizes input data. The \textit{Radiomics Pre-Checks} step provides exploratory analyses to guide the selection of feature-extraction settings.}
  \label{fig:datamanager}
\end{figure}

The \textit{DataManager} module provides graphical access to the \textit{DataManager} Python class. It provides a structured environment for organizing and managing imaging datasets prior to cohort-level analyses. Users can register imaging studies and their associated segmentation masks and extract relevant image metadata. The module also provides exploratory analyses of the dataset, including voxel intensity distribution statistics and histogram visualizations to help users select the appropriate preprocessing and feature extraction settings required in the \textit{BatchExtractor} step. An overview of the \textit{DataManager} module is presented in Figure \ref{fig:datamanager}.

\subsubsection{BatchExtractor}
\label{sec:medapp_batchExtractor}
\begin{figure}[!htbp]
  \centering
  \includegraphics[scale=0.302]{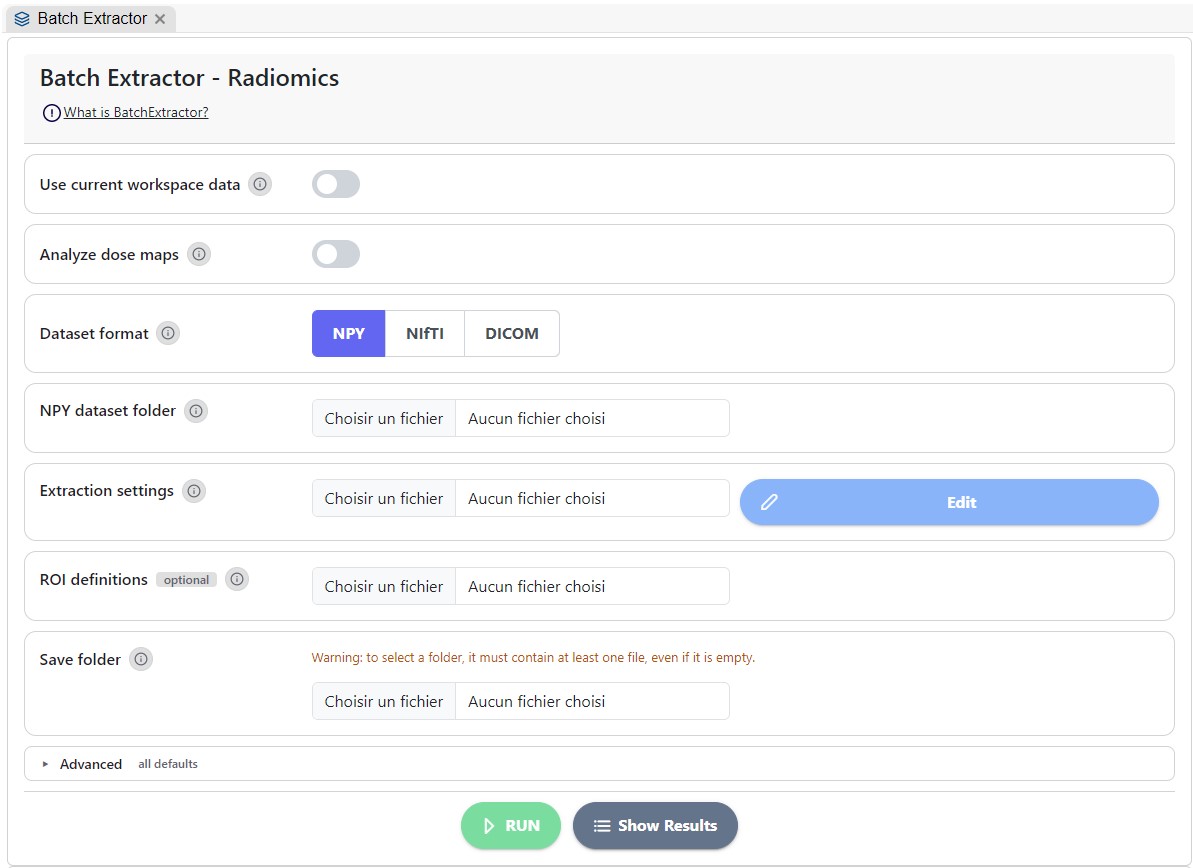}
   \caption{\textbf{Overview of the \textit{BatchExtractor} module.} Extraction settings are configured per imaging modality for cohort-level radiomic feature extraction.}
  \label{fig:batchextractor}
\end{figure}

The \textit{BatchExtractor} module extends the extraction capabilities of the \textit{Single-Scan Extraction} module to full imaging cohorts. Once a dataset has been registered through the \textit{DataManager}, the \textit{BatchExtractor} allows users to configure extraction parameters, for each imaging modality, at the cohort level and run parallelized feature extraction across all subjects. Similar to the Python Package, features are extracted independently for each scan, but a final CSV file is generated to group all extracted features, representing the entire dataset, producing an output file compatible with the \textit{Learning} module. An overview of the \textit{BatchExtractor} module is presented in Figure \ref{fig:batchextractor}.

\paragraph{BatchExtractor - Feature Categorization:}
\label{sec:medapp_feats_cat}
To support model training based on feature families, the \textit{MEDiml} application includes a feature categorization step, within the \textit{BatchExtractor} module, that splits an extracted radiomics table into separate subsets, one per feature family. The step runs after radiomic features have been extracted from a dataset with the batch extractor, and before the machine learning analysis. Each feature is assigned to a family from the prefix of its standardized name, as listed in the definitions file that accompanies the table. Morphological features carry the prefix \textit{morph\_}, intensity-based features carry \textit{locInt\_}, \textit{stats\_}, \textit{intHist\_} or \textit{intVolHist\_}; and texture features come from the GLCM, GLRLM, GLSZM, NGLDM, GLDZM and NGTDM matrices. Features matching none of these prefixes are placed in an “unknown” set rather than discarded. Each family (morphological, intensity, texture, or unknown) is saved as its own table and definitions file, organized by imaging modality. The tables keep the naming convention the \textit{Learning} module expects, so each family can then be analyzed on its own.

\begin{figure}[htbp]
  \centering
  \includegraphics[scale=0.342]{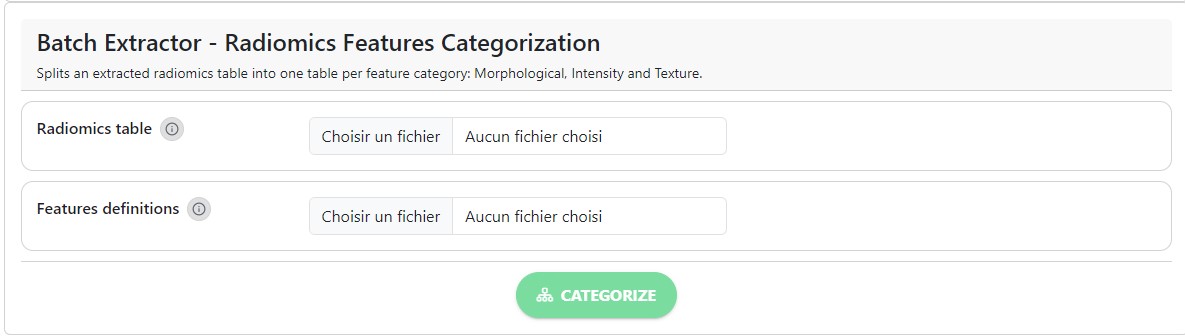}
  \caption{\textbf{Feature Categorization card in the \textit{BatchExtractor} module}. Extracted features are grouped by feature family for subsequent analyses.}
  \label{fig:feat_categ}
\end{figure}

\subsubsection{Learning}
\label{sec:medapp_learning}
The \textit{Learning} module implements the complete \textit{MEDiml} machine learning workflow and operates on features previously extracted through the BatchExtractor. Similar to the \textit{Single-Scan Extraction} module, the \textit{Learning} module is represented as a node-based scene in which users visually assemble and configure the analytical pipeline (Figure \ref{fig:ml_scene}).

The available node library spans all stages of radiomics-based predictive modeling, including:

\begin{itemize} 
 \item \textbf{Split}: dataset partitioning and validation strategy definition;
 \item \textbf{Design}: experimental design and radiomic complexity level selection;
 \item \textbf{Data}: loading of extracted feature sets;
 \item \textbf{Cleaning}: removal of invariant, duplicate, or otherwise non-informative features;
 \item \textbf{Normalization}: feature standardization and harmonization via ComBat \cite{combat};
 \item \textbf{Feature Reduction}: feature selection using the False Discovery Avoidance (FDA) methodology \cite{fda_21};
 \item \textbf{Radiomics Learner}: model training and evaluation;
 \item \textbf{Analyze}: performance assessment, statistical significance testing, AUC \cite{fawcett_roc_2006} analysis, and feature importance visualization.
\end{itemize}

Multiple pipelines can be created and executed simultaneously within the same scene, enabling direct comparison of different radiomic complexity levels, datasets, or imaging modalities. Results are displayed interactively within the application and can be exported for reporting, publication, or further downstream analyses.

Through this workflow-centric design, the \textit{Learning} module transforms complex radiomics and machine learning procedures into an accessible visual environment while maintaining methodological rigor and full compatibility with the underlying \textit{MEDiml} Python framework. 

\section{System Design}
\label{sec:sys_design}
\textit{MEDiml}’s system design connects graphical workflow construction with Python execution, so that researchers can inspect and extend the code corresponding to an analysis configured through the application. This integration requires the interface, backend (execution layer) and code-generation mechanism to preserve the configured analysis settings. The remainder of this section is structured as follows: Subsections \ref{sec:arch_overview} and \ref{sec:Technology_Stack} describe the architecture and technology stack. Subsection \ref{sec:code_gen} explains code generation, and Subsections \ref{sec:gpu_filtering} and \ref{sec:dosiomics_support} describe accelerated filtering \cite{buck_cuda_2007, klockner_pycuda_2012} and dosiomics \cite{liang_dosiomics_2019} support. Finally, Subsection \ref{sec:ibsi_compliance} describes IBSI \cite{ibsi_1, ibsi_2} compliance.

\subsection{Architecture Overview}
\label{sec:arch_overview}
\textit{MEDiml} was designed to bridge the gap between clinical researchers and computational scientists by providing both a graphical and a programmatic interface to the same radiomics framework. As illustrated in Figure \ref{fig:introductory}, the platform consists of two integrated components: a desktop graphical user interface (GUI) for users without programming experience and a Python package for users requiring direct access to computational workflows. Rather than functioning as independent software products, both components use the same computational backend, ensuring that all analyses are executed using the same implementation.

This architecture provides three major advantages. First, it guarantees methodological consistency between graphical and programmatic workflows. Second, it promotes reproducibility by ensuring that identical operations produce identical results regardless of the chosen interface. Third, it facilitates collaboration between multidisciplinary teams by enabling clinical researchers to design experiments visually while allowing computational researchers to access, reproduce, and extend the same workflows programmatically. All information about the software is available here: \href{https://www.mediml.app/}{mediml.app}.

\subsection{Technology Stack}
\label{sec:Technology_Stack}
\textit{MEDiml} combines modern software technologies selected to maximize scalability and usability (Figure \ref{fig:introductory}, Block 3, panels B and D).

The desktop application is built using Electron \cite{electron}, an open-source framework that packages web applications as native desktop software. This approach enables \textit{MEDiml} to run on Windows, macOS, and Linux from a single codebase while maintaining a consistent user experience across operating systems.

The graphical interface is implemented using React \cite{react}, a component-based JavaScript framework that facilitates the development of responsive and interactive user interfaces. React provides the modular structure required to organize the numerous components involved in radiomics analyses while maintaining smooth interaction during workflow configuration and execution.

To support visual workflow design, \textit{MEDiml} integrates ReactFlow \cite{reactflow}, a specialized library for graph-based editors. ReactFlow enables users to construct radiomics pipelines by connecting nodes representing individual computational tasks, such as image loading, preprocessing, feature extraction, feature selection, machine learning, and result analysis.

Communication between the graphical interface and the computational backend is managed through Go, also known as Golang \cite{golang}, selected because of its performance characteristics and support for parallel execution. Within \textit{MEDiml}, the Go backend translates user-defined workflows into structured instructions, coordinates execution requests to the Python-based computational backend, and returns the resulting outputs to the interface for visualization and interpretation.

Finally, the \textit{MEDiml} Python package builds upon established scientific computing and machine learning libraries, including NumPy \cite{harris_numpy_2020} for numerical computation, Pandas \cite{pandas_2020} for tabular data management, SciPy \cite{virtanen_scipy_2020} for statistical analysis, Scikit-learn \cite{pedregosa_sklearn_2011} and PyCaret \cite{ali_pycaret_2024} for machine learning, and Ray \cite{moritz_ray_2018} for parallelized processing. By leveraging these libraries, \textit{MEDiml} provides a scalable and efficient framework for radiomics analyses.

\subsection{Code Generation}
\label{sec:code_gen}
A software contribution of \textit{MEDiml} is the integration between its graphical interface and Python package. Every operation available through the user interface is implemented using the underlying Python framework. When a user constructs an analysis pipeline graphically, including the preprocessing parameters, feature extraction settings, feature selection options, machine learning models, and results analysis schemes, the workflow definition is translated into structured instructions and executed by the Python backend (Figure \ref{fig:introductory}, Block 1).

\begin{figure}[!htbp]
  \centering
  \includegraphics[scale=0.402]{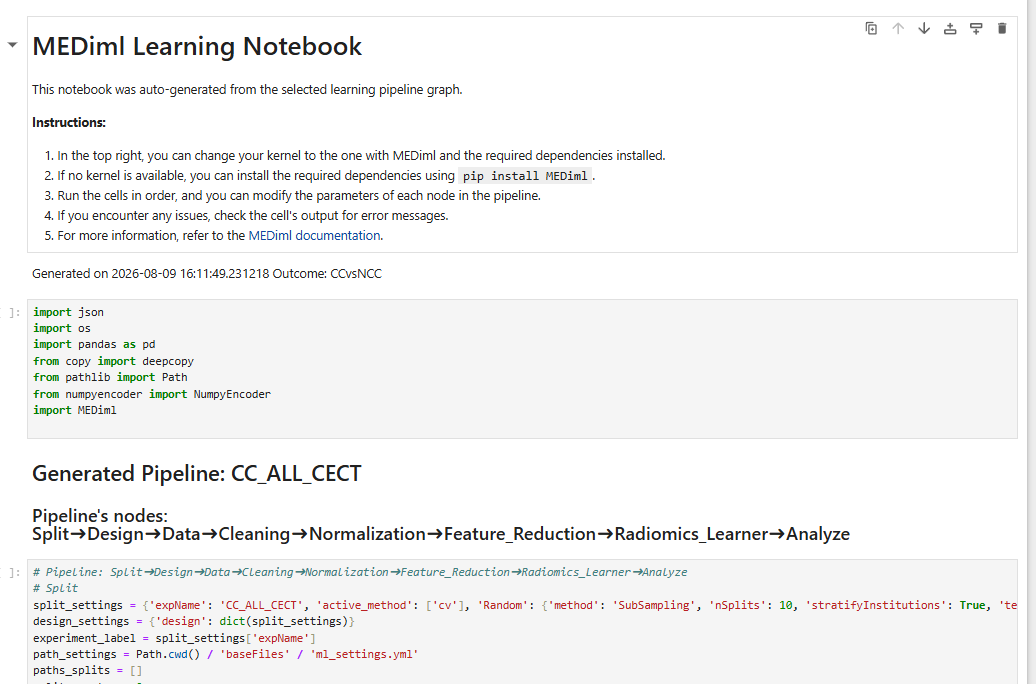}
  \caption{\textbf{Example of a generated Jupyter Notebook.} The exported notebook contains executable Python code corresponding to a graphically configured workflow \cite{kluyver_jupyter_2016}.}
  \label{fig:jupyter_notebook}
\end{figure}

Because both interfaces rely on the same computational engine, graphical and programmatic workflows are functionally equivalent. Following successful execution, users may generate executable code for any pipeline directly from the interface. This functionality is implemented through the \textit{NotebookBuilder} class, which traverses the workflow graph and automatically converts each node and its associated configuration into Python code. The generated workflow is exported as a Jupyter Notebook \cite{kluyver_jupyter_2016}, an interactive web-based environment that combines executable Python code, documentation, visualizations, and analysis outputs within a single page. An example is shown in Figure \ref{fig:jupyter_notebook}.

By generating notebooks rather than standalone scripts, \textit{MEDiml} enables users to inspect, modify, extend, and re-execute analyses interactively while preserving the exact experimental configuration defined through the graphical interface. This architecture eliminates the need to manually recreate experiments across environments, facilitates collaboration between clinical researchers and computational scientists, and promotes reproducibility.

\subsection{GPU-Accelerated Filtering}
\label{sec:gpu_filtering}
Radiomic analyses involving nonlinear image filters are computationally intensive, particularly when local texture measurements must be computed repeatedly across large three-dimensional image volumes. To address this challenge, \textit{MEDiml} implements a GPU-accelerated filtering framework specifically designed for texture-based image transformation. Computing textural filter maps on the GPU is an additional software contribution supporting \textit{MEDiml}’s feature-extraction workflow.

To maximize computational performance, the filtering kernels and the computation of GLCM features used to generate textural filter maps are implemented entirely in C++, and Compute Unified Device Architecture (CUDA) \cite{buck_cuda_2007} is used for parallel processing on GPUs. Computationally intensive mathematical operations, including functions such as \textit{sqrtf} and other CUDA-optimized functions, are executed directly on the GPU, enabling thousands of voxel-level texture computations to be performed simultaneously. This parallelization substantially reduces processing times compared with traditional implementations running on central processing units (CPUs).

CUDA \cite{buck_cuda_2007} kernels are compiled and executed through \textit{PyCUDA} \cite{klockner_pycuda_2012}, which serves as the interface between Python and the GPU implementation. \textit{PyCUDA} is responsible for kernel compilation, device memory allocation and transfer, execution management, and runtime error handling, allowing the high-performance C++/CUDA filtering engine to be seamlessly integrated into the \textit{MEDiml} Python framework. As a result, computationally demanding textural filtering operations can be incorporated into routine radiomics workflows while maintaining practical execution times, even for large imaging cohorts and high-dimensional feature extraction pipelines.

\subsection{Dosiomics Support}
\label{sec:dosiomics_support}
In addition to conventional radiomics analyses, \textit{MEDiml} supports dosiomics \cite{liang_dosiomics_2019}, an emerging field that applies quantitative image analysis techniques to radiotherapy dose distributions. Because dose maps can be represented as volumetric images, \textit{MEDiml} can process them using the same workflow employed for other imaging modalities, including preprocessing, feature extraction, feature reduction, predictive modeling, and result interpretation. This allows researchers to investigate the relationship between spatial dose distribution patterns and clinical outcomes.

Unlike conventional radiomics, dosiomics \cite{liang_dosiomics_2019} extends beyond the feature definitions established by the Image Biomarker Standardization Initiative (IBSI) \cite{ibsi_1, ibsi_2}. Consequently, \textit{MEDiml} supplements IBSI-compliant radiomic features with a dedicated collection of dose-specific biomarkers commonly used in radiotherapy research and treatment plan evaluation. These include dose-volume histogram (DVH)-derived metrics such as $D_{\mathrm{max}}$, $D_{\mathrm{mean}}$, $D_{\mathrm{2\%}}$, $D_{\mathrm{95\%}}$, and $D_{\mathrm{98\%}}$, as well as treatment-planning indices including the Conformity Index (CI), Gradient Index (GI), Heterogeneity Index (HI), Selectivity, and Coverage. Additional dose-based measurements, such as V$_x$ metrics, where x indicates the dose threshold, i.e., the fraction of volume receiving at least $x$ Gy of planned dose, and isocenter-related parameters, are also supported. These biomarkers characterize dose coverage, conformity, gradient fall-off, and dose heterogeneity within the target volume, providing complementary information that is not captured by traditional image-derived radiomic features.

By combining standardized radiomic biomarkers with dose-specific descriptors, \textit{MEDiml} provides a comprehensive framework for dosiomics \cite{liang_dosiomics_2019} studies and enables the development of predictive models that incorporate both anatomical and dosimetric information. This facilitates the investigation of treatment response, toxicity prediction, and outcome modeling within modern radiation oncology workflows. Integrating dose-specific descriptors into the shared radiomics workflow is an additional software contribution of \textit{MEDiml}.

\subsection{IBSI Compliance}
\label{sec:ibsi_compliance}
\textit{MEDiml} was calibrated in accordance with IBSI chapter 1 \cite{ibsi_manu_1_25} and 2 \cite{ibsi_manu_2} through direct participation in the IBSI benchmarking and validation process. Specifically, the \textit{MEDiml} team participated in all validation phases of both chapters by applying the software to the reference datasets and phantom objects provided by the IBSI team. This process included image preprocessing, image filtering, and feature extraction, followed by submission of the resulting feature values to the IBSI coordinators for comparison against results obtained by other participating software implementations.

Through its participation in the IBSI \cite{ibsi_1, ibsi_2} standardization effort, \textit{MEDiml} adheres to international radiomic standards, promotes the computation of reproducible biomarkers and facilitates the comparison of results across datasets, institutions, and software implementations. This standardization is essential for improving the robustness and clinical translatability of radiomics-based predictive models.

\subsection{How to install \textit{MEDiml}}
\label{sec:install}
\textit{MEDiml} can be installed as a Python package, for programmatic analyses, or as a desktop application, for analyses through the graphical interface. Subsection \ref{sec:install_pkg} describes the installation of the Python package, and Subsection \ref{sec:install_app} describes the installation of the application.

\subsubsection{Python package}
\label{sec:install_pkg}
The \textit{MEDiml} package is distributed through the Python Package Index (PyPI) and requires Python 3.8 or later. The latest stable version can be installed with:
\begin{verbatim}
pip install MEDiml
\end{verbatim}
Users can also install a specific version, for example, to reproduce a previous analysis, by replacing \texttt{x.x.x} with the desired version number (see all releases in \href{https://github.com/MEDomicsLab/MEDiml-pkg/releases}{MEDiml/releases}):
\begin{verbatim}
pip install MEDiml==x.x.x
\end{verbatim}
Finally, the most recent changes can be installed directly from the development branch of the GitHub repository. This version includes the latest features and fixes, but is less stable than the released versions:
\begin{verbatim}
pip install git+https://github.com/MEDomicsLab/MEDiml-pkg@dev
\end{verbatim}

\subsubsection{Desktop application}
\label{sec:install_app}
To install the \textit{MEDiml} application, users download the release file corresponding to their operating system (Windows, macOS or Linux) and open it to automatically install the application; no further instructions are necessary. When the application is opened, a setup window appears: users click \textit{Install} and wait until all requirements are installed before closing the application. If a problem occurs, the \href{https://medomicslab.gitbook.io/mediml-app-docs/troubleshooting}{troubleshooting page} of the documentation describes how to diagnose and fix common issues.

\section{\textit{MEDiml} Vs. Existing Radiomics Tools}
\label{sec:mediml_vs_rest}
To place \textit{MEDiml} in context, Table \ref{tab:comparison} compares the analytical capabilities and practical access of existing radiomics tools. These tools typically prioritize either feature extraction, graphical usability, or predictive modeling, but rarely all three simultaneously. Several platforms provide robust radiomics engines and varying degrees of IBSI \cite{ibsi_1, ibsi_2} compliance, although partial compliance often indicates that only part of the IBSI recommendations or validation phases has been implemented. In many cases, predictive modeling is either absent or limited to isolated functionalities, requiring users to export features to external software for feature selection, model development, or statistical analysis. As a result, researchers frequently rely on multiple software packages to complete a single radiomics study.

\begin{table}[ht]
\centering
\footnotesize
\renewcommand{\arraystretch}{1.3}
\caption{\textbf{Comparison of \textit{MEDiml} with commonly used radiomics tools.} The tools are compared by interface type, programming language, compliance with Image Biomarker Standardization Initiative (IBSI) guidelines, radiomics and machine learning capabilities, execution environment, code generation, and source-code availability. A check mark indicates that a capability is available; a cross indicates that it is unavailable. ``Partial'' denotes limited compliance with the IBSI, and N/A means not applicable.}\medskip
\begin{tabularx}{\textwidth}{|>{\raggedright\arraybackslash}p{0.13\textwidth}|>{\raggedright\arraybackslash}X|>{\raggedright\arraybackslash}X|>{\raggedright\arraybackslash}X|*{5}{>{\centering\arraybackslash}X|}}
\hline
\textbf{Package/ Software} & \textbf{Type} & \textbf{Language} & \textbf{IBSI compliance} \cite{ibsi_1, ibsi_2} & \textbf{Own radiomics engine} & \textbf{Machine learning integration} & \textbf{Local execution} & \textbf{Code generation} & \textbf{Open-source} \\ \hline
PyRadiomics \cite{vangriethuysen_pyradiomics_2017} & Library                         & Python             & Partial         & \cmark & \xmark  & \cmark & N/A    & \cmark \\ \hline
3D Slicer (SlicerRadiomics) \cite{fedorov_slicer_2012, vangriethuysen_pyradiomics_2017} & GUI (Plugin)                    & Python             & Via PyRadiomics & \xmark & \xmark  & \cmark & \xmark & \cmark \\ \hline
QuantImage V2 \cite{quantimage_v2} & GUI (web)                       & Python, JavaScript & Via PyRadiomics & \xmark & \cmark  & \xmark & \xmark & \cmark \\ \hline
MIRP \cite{zwanenburg_mirp_2024} & Library                         & Python             & IBSI 1 \& 2     & \cmark & \xmark  & \cmark & N/A    & \cmark \\ \hline
LIFEx \cite{nioche_lifex_2018} & GUI (Freeware)                  & Java               & Partial         & \xmark & \cmark  & \cmark & \xmark & Free / Closed Source \\ \hline
CERR \cite{cerr_20} & Library and GUI                 & MATLAB             & IBSI 1 \& 2     & \cmark & Partial & \cmark & \xmark & \cmark \\ \hline
MITK \cite{wolf_mitk_2005} & Library and GUI (desktop app)   & C++                & Partial         & \cmark & Partial & \cmark & \xmark & \cmark \\ \hline
PySERA \cite{salmanpour_pysera_2026} & Library                         & Python             & IBSI 1 \& 2     & \cmark & \cmark  & \cmark & N/A    & \cmark \\ \hline
\textbf{MEDiml} & \textbf{Library and GUI (desktop app)} & \textbf{Python, JavaScript} & \textbf{IBSI 1 \& 2} & \cmark & \cmark & \cmark & \cmark & \cmark \\ \hline
\end{tabularx}
\label{tab:comparison}
\end{table}

\textit{MEDiml} was designed to address these limitations through a unified framework that combines an IBSI-compliant \cite{ibsi_1, ibsi_2} radiomics engine, integrated machine learning, a graphical user interface, and a Python package. Unlike existing solutions, the graphical and programmatic environments are directly connected through automatic code generation, enabling workflows designed visually to be reproduced and extended in Python without reimplementation. Furthermore, \textit{MEDiml} introduces a complexity-driven approach \cite{loutfi_complexity_2024} that emphasizes the identification of compact, interpretable radiomic signatures rather than increasingly large feature sets, facilitating reproducible analyses and simplifying the integration of imaging biomarkers with other data modalities in multimodal predictive modeling.

\section{Use Cases: \textit{MEDiml} Evaluation}
\label{sec:use_cases}
To evaluate \textit{MEDiml}’s complexity-driven approach \cite{loutfi_complexity_2024}, we compare the predictive performance of radiomic feature sets across five oncological imaging cohorts. The prediction tasks cover histological subtype, tumor grade and isocitrate dehydrogenase 1 (IDH1) mutation status. The remainder of this section is structured as follows: Subsections \ref{sec:cohorts} to \ref{sec:stat_analysis} describe the cohorts, feature-extraction settings and statistical analysis, and Subsection \ref{sec:results} presents the results and their interpretation.

\subsection{Cohorts}
\label{sec:cohorts}
Five distinct imaging cohorts of cancer patients, including non-small cell lung cancer (NSCLC), low-grade glioma (LGG), meningioma and two imaging cohorts of renal cell carcinoma (RCC) were collected. The NSCLC dataset was collected from the study by Primakov \textit{et al.} study \cite{nsclc_study}. It includes data from three institutions: MAASTRO \cite{nsclc_dataset_10}; Stanford \cite{nsclc_dataset_11}, available on The Cancer Imaging Archive (TCIA) \cite{TCIA_12}; and the University of California San Francisco (UCSF), which is not public. For LGG, part of the data was collected from The Cancer Genome Atlas (TCGA) \cite{lgg_tcga_13}, while the rest was provided by Yu \textit{et al.} and Li \textit{et al.} studies \cite{lgg_dataset_14, lgg_dataset_15}, respectively. Meningioma cohort was provided by the Wu \textit{et al.} \cite{wu_presenting_2018}, the Vasudevan \textit{et al.} \cite{vasudevan_comprehensive_2018}, the Gennatas \textit{et al.} \cite{gennatas_preoperative_2018} and the Morin \textit{et al.} \cite{meningioma_dataset_16} studies. MRI-based RCC cohort was provided by the Xi \textit{et al.} study \cite{rcc_dataset_17}, and TCGA data \cite{dataset_rcc_18} was included. Finally, the contrast-enhanced computed tomography (CECT)-based RCC dataset was provided by CIUSSS de l'Estrie Centre hospitalier universitaire de Sherbrooke (CIUSSSE-CHUS). Institutional review approval was given for the use of this dataset in this retrospective study. The cohorts are summarized in table \ref{tab:dataset} (patient characteristics are provided in supplementary note 3 of \cite{loutfi_complexity_2024}).

\begin{table}[ht]
    \scriptsize
    \caption{\textbf{Summary of the five study cohorts.}}\medskip
    \label{tab:dataset}
    \begin{tabular}{|llllll|}
    \hline
    \multicolumn{1}{|l|}{Cohorts}                                   & \multicolumn{1}{c|}{\begin{tabular}[c]{@{}c@{}}NSCLC\\ (n=506)\end{tabular}}                                                                            & \multicolumn{1}{c|}{\begin{tabular}[c]{@{}c@{}}LGG\\ (n=329)\end{tabular}}                                                        & \multicolumn{1}{c|}{\begin{tabular}[c]{@{}c@{}}Meningioma\\ (n=344)\end{tabular}}                                                                            & \multicolumn{1}{c|}{\begin{tabular}[c]{@{}c@{}}RCC\\ (n=599)\end{tabular}}                                                                                                       & \multicolumn{1}{c|}{\begin{tabular}[c]{@{}c@{}}RCC\\ (n=326)\end{tabular}}                                                             \\ \hline
    \multicolumn{1}{|l|}{Institutions}                              & \multicolumn{1}{l|}{\begin{tabular}[c]{@{}l@{}}MAASTRO (n=207)\cite{nsclc_study, nsclc_dataset_10}\\ UCSF (n=163) \cite{nsclc_study}\\ Stanford (n=136)\cite{nsclc_study, nsclc_dataset_11}\end{tabular}}                                & \multicolumn{1}{l|}{\begin{tabular}[c]{@{}l@{}}TCGA (n=103)\cite{lgg_tcga_13}\\ Huashan (n=226)\cite{lgg_dataset_14}\\ \cite{lgg_dataset_15}\end{tabular}}                          & \multicolumn{1}{l|}{\begin{tabular}[c]{@{}l@{}}UCSF (n=257)\cite{wu_presenting_2018}\\\cite{vasudevan_comprehensive_2018, gennatas_preoperative_2018, meningioma_dataset_16}\\ PM (n=87)\cite{meningioma_dataset_16}\end{tabular}}                                                              & \multicolumn{1}{l|}{\begin{tabular}[c]{@{}l@{}}Penn (n=439)\cite{rcc_dataset_17}\\ Mayo (n=53) \cite{rcc_dataset_17}\\ TCGA (n=54)\cite{dataset_rcc_18}\\ HPH (n=30)\cite{rcc_dataset_17}\\ XYSH (n=23)\cite{rcc_dataset_17}\end{tabular}}                        & \begin{tabular}[c]{@{}l@{}}CIUSSSE-CHUS\\ (n=326)\end{tabular}                                                                         \\ \hline
    \multicolumn{1}{|l|}{Imaging modality}                          & \multicolumn{1}{c|}{CT}                                                                                                                                 & \multicolumn{1}{c|}{MRI-T2F}                                                                                                      & \multicolumn{1}{c|}{MRI-T1CE}                                                                                                                                & \multicolumn{1}{c|}{MRI-T2WI}                                                                                                                                                    & \multicolumn{1}{c|}{CECT}                                                                                                              \\ \hline
    \multicolumn{1}{|l|}{Clinical endpoint}                         & \multicolumn{1}{l|}{\begin{tabular}[c]{@{}l@{}}Histological subtype:\\ - Adenocarcinoma (n=240)\\ - Other (n=266)\end{tabular}}                         & \multicolumn{1}{l|}{\begin{tabular}[c]{@{}l@{}}IDH1 mutation:\\ - Yes (n=239)\\ - No (n=90)\end{tabular}}                         & \multicolumn{1}{l|}{\begin{tabular}[c]{@{}l@{}}Pathological tumor \\grade*:\\ - Grade 1 (n=197)\\ - Grade 2 \& 3 (n=147)\end{tabular}}                         & \multicolumn{1}{l|}{\begin{tabular}[c]{@{}l@{}}Subtype discrimination:\\ - Papillary (n=158)\\ - Clear Cell (n=441)\end{tabular}}                                                & \begin{tabular}[c]{@{}l@{}}Subtype discrimination\\ - Non-Clear Cell (n=79)\\ - Clear Cell (n=247)\end{tabular}                        \\ \hline
    \multicolumn{6}{|l|}{\begin{tabular}[c]{@{}l@{}}NSCLC: non-small cell lung cancer. \\CT: Computed tomography. MRI: Magnetic resonance imaging. T2F: T2-weighted/FLAIR (Fluid attenuated inversion recovery).\\T1CE: T1-weighted contrast-enhanced. T2WI: T2-weighted image. \\ CECT: Contrast-enhanced computed tomography. UCSF: University of California, San Francisco. PM: Princess Margaret Cancer Centre. \\ CIUSSSE-CHUS: Centre intégré universitaire de santé et de services sociaux de l’Estrie-Centre hospitalier universitaire de Sherbrooke. \\ Penn: Hospital of the University of Pennsylvania. Mayo: Mayo Clinic. HPH: Hunan Provincial People's Hospital. \\ TCGA: The Cancer Genome Atlas. XYSH: Xiangya Second Hospital of Central South University. \\ *For binary prediction of pathological grade, Grade 1 is considered “Low” (0), and Grade 2 and 3 is considered “High” (1).\end{tabular}} \\ \hline
    \end{tabular}
    \end{table}

\subsection{Radiomic Features and Complexity Levels}
\label{sec:feats_and_levels}

Recall that a radiomic complexity level is a candidate feature set defined by the feature families and extraction operations it includes (see more in Section~\ref{sec:mediml_workflow}). In this evaluation, models are compared across sets derived from the original image and from filtered images, so that the contribution of more elaborate extraction operations can be assessed for each clinical task.

For each dataset, a total of 170 features were extracted from the original images, in accordance with IBSI definitions \cite{ibsi_1}: 27 shape-based; 49 intensity-based (1 local intensity, 18 statistical, 23 intensity histogram and 7 intensity-volume histogram); and 94 texture-based, each extracted using six different combinations of parameters, for images with physical units (CT and CECT) and three different combinations of parameters for the rest (MRI and filtered images). For filtered images, shape and intensity-based statistical features were excluded as they are not meaningful for arbitrary intensity scales, resulting in a total of 124 features. Six LF were used, including mean, Laplacian-of-Gaussian, Laws, Gabor, wavelets (high-pass and low-pass), all in accordance with IBSI definitions \cite{ibsi_2}. We also used nonlinear filtering by moving a predefined three-dimensional cubic window over the voxels of the image and calculating texture feature values from the Gray Level Co-occurrence Matrix (GLCM) \cite{sebastian_glcm_2012} family at each position (hereafter referred to as “textural filtering”). This resulted in 25 textural filters (TF) corresponding to the 25 features defined in the GLCM family. To our knowledge, no reference values currently exist for such filters, but similar definitions are found in the works of Mayerhoefer \textit{et al.} \cite{intro_radiomics_19}, and of Deasy \textit{et al.} \cite{cerr_20}. Consequently, a total of 18,112 features were extracted for cohorts with physical units, and 17,830 features for cohorts with arbitrary ones. Image processing and extraction settings are provided in supplementary note 1 of \cite{loutfi_complexity_2024}.

The aforementioned categories of radiomic features are henceforth designated as complexity levels. Therefore, our investigation focuses on five levels of complexity. The first three levels include features extracted from the original image intensities: Morphological (“M”) features, Intensity-based (“I”) features and Texture-based (“T”) features. The fourth (“LF”) and fifth (“TF”) levels are features extracted after linear and textural filtering, respectively. To increase complexity, features were combined prior to predictive modeling, giving rise to the final sequence of complexity levels: “M”, “I”, “M+I” (MI), “T”, “M+I+T” (MIT), “LF”, “M+I+T+LF” (MITLF), “TF”, and “M+I+T+LF+TF” (MITLFTF).

\subsection{Statistical analysis}
\label{sec:stat_analysis}
The area under the curve (AUC) \cite{fawcett_roc_2006} metric was employed for evaluation. To address the issues encountered with small datasets, including overfitting, noise, outliers and sampling bias, which can render the learned model ineffective \cite{issue_small_data_22}, model predictions over all 10 test folds of the cross-validation \cite{arlot_cv_2010} were aggregated into a single receiver operating characteristic (ROC) curve, mimicking the behavior of leave-one-out cross-validation. While this approach sacrifices the model’s variance, it is known for minimizing bias and offering reliable estimates \cite{cv_review_23}. A DeLong test \cite{delong_1988, delong_24} was used to determine whether models were statistically different, and the difference was considered significant with $\text{p-value}<.05$.

\subsection{Results}
\label{sec:results}
This subsection presents the results of the use-case evaluation. Subsection \ref{sec:level_identification} reports, for each cohort, the optimal complexity level identified by \textit{MEDiml} and the features with the highest importance at that level, as well as the performance of the filter-based levels. Subsection \ref{sec:in_depth_analysis} then presents an in-depth analysis of the optimal level identified for the CECT-based RCC cohort, showing how this level can guide the refinement of feature extraction settings.

\subsubsection{Optimal complexity level identification and explanation}
\label{sec:level_identification}
\begin{figure}
  \centering
  \includegraphics[scale=0.172]{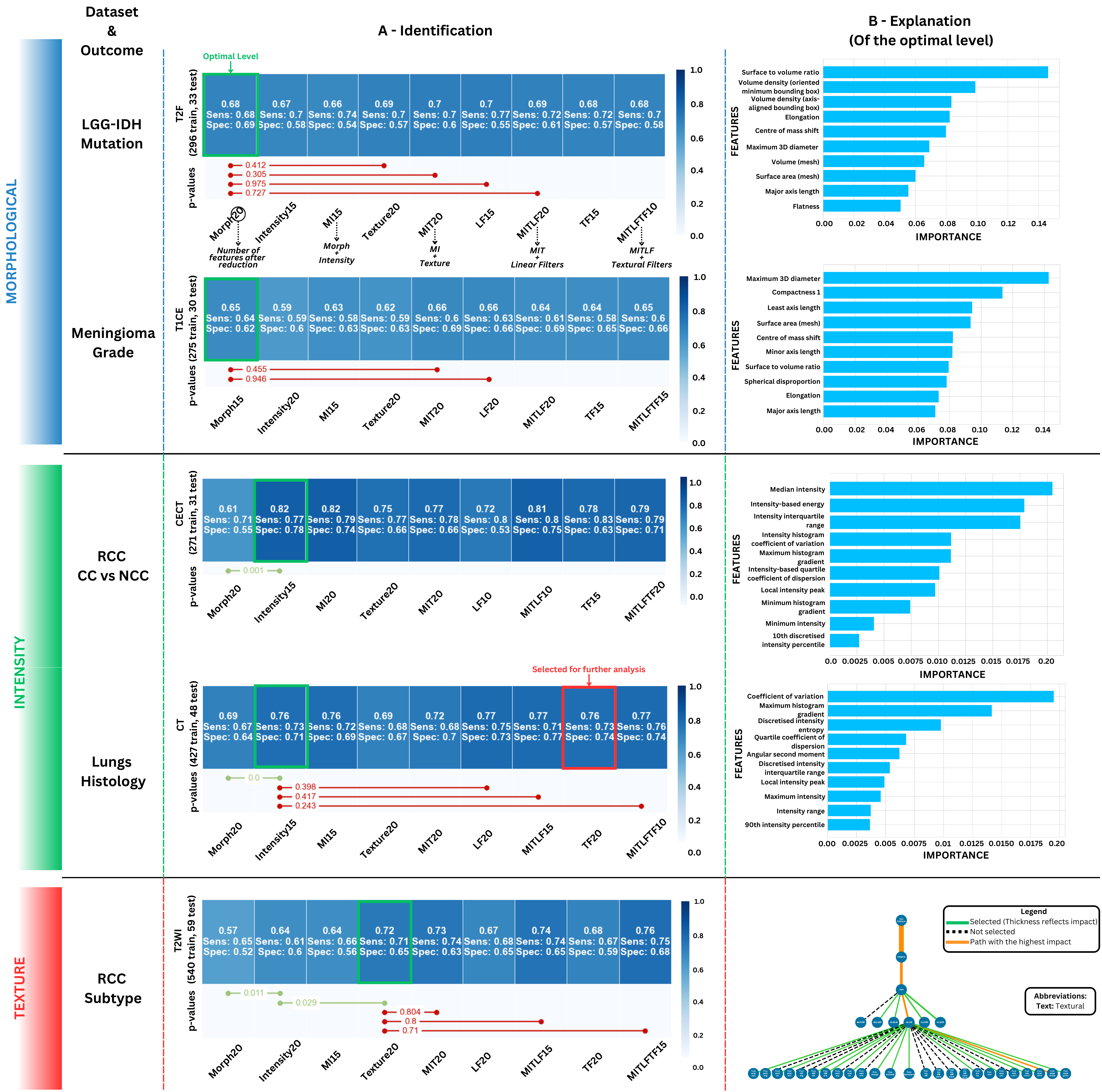}
 \caption{\textbf{Selected radiomic complexity levels and feature importance across the five cohorts.} Levels are identified as blue (morphological), green (intensity), and red (texture). A. Comparison of heatmaps showing aggregated AUC \cite{fawcett_roc_2006} across splits for various radiomics signatures. Each column indicates a complexity level. The numerical suffixes in column names indicate the number of features retained after feature set reduction. In the bottom rows, green lines indicate significant improvements ($\text{p-value}<.05$) and red lines indicate non-significant improvements. Green boxes point to the identified optimal levels and the red one is selected for further discussion in the supplementary material of \cite{loutfi_complexity_2024}. B: Explanation section displaying a histogram of feature importance or importance tree, highlighting features with high importance at the identified optimal level.}
  \label{fig:results}
\end{figure}

Model performance was assessed using the aggregated AUC \cite{fawcett_roc_2006} value across cross-validation \cite{arlot_cv_2010} splits as a base metric, and the mean feature importance across splits was used to identify highly predictive features. All results are depicted in Figure \ref{fig:results}.

\paragraph{Morphological features as the optimal complexity level:} 
In the Meningioma cohort, the model solely utilizing morphological features displayed an AUC of 0.65 (95\% CI: 0.59, 0.72), specificity of 0.62, and sensitivity of 0.64. Models based on other complexity levels displayed varying degrees of performance, such as intensity-based (AUC 0.59), MI-based (AUC 0.63), Texture-based (AUC 0.62), and TF-based (AUC 0.64). MIT-based and MITLF-based models had the highest specificity (0.69) but a lower sensitivity (0.6 and 0.61, respectively). MIT-based and LF-based models exhibited the highest AUC (0.66), yet were not significantly different from the morphological features-based model (p = .46 and .95 respectively). Therefore, the morphology-based features based model was selected as optimal. The maximum three-dimensional diameter feature had the highest mean importance in the model. Similarly, for the prediction of IDH1 mutation in LGG, morphological features were sufficient to obtain the best performance, with an AUC of 0.68 (95\% CI: 0.60, 0.75), specificity of 0.69, and sensitivity of 0.68. The surface-to-volume ratio feature had the highest mean importance.

\paragraph{Intensity features as the optimal complexity level:}
For histological subtype classification of NSCLC, the intensity features model achieved an AUC of 0.76 (95\% CI: 0.71, 0.80), specificity of 0.71, and sensitivity of 0.73. Models based on LF, MITLF and MITLFTF demonstrated higher AUC values (0.77), yet their p-values (.40, .42 and .24, respectively) did not indicate statistical significance with respect to the intensity features model. The coefficient of variation had the highest mean feature importance. Similarly, for the clear and non-clear RCC classification based on CECT, the model based on intensity features recorded the highest AUC of 0.82 (95\% CI: 0.76, 0.88), with a specificity of 0.78 and a sensitivity of 0.77. Median intensity had the highest mean feature importance.

\paragraph{Texture features as the optimal complexity level:}
Texture-based model proved improvement in performance from morphological-based (p=.01) and intensity-based (p=.03) models, for the subtype classification of RCC based on MRI-T2WI, achieving an AUC of 0.72 (95\% CI: 0.68, 0.77), specificity of 0.65, and sensitivity of 0.71. Within the texture feature families, the GLCM \cite{sebastian_glcm_2012} feature family demonstrated the highest mean feature importance, with cluster shade having the highest importance.

\paragraph{Linear filters and textural filters:}
Although the LF-based models were never selected as optimal, they consistently demonstrated high AUC values across cohorts. For example, in the classification of NSCLC subtypes, the LF-based model achieved an AUC of 0.77 (95\% CI: 0.72, 0.80), indicating their potential utility in radiomics analyses.

\begin{figure}
  \centering
  \includegraphics[scale=0.4]{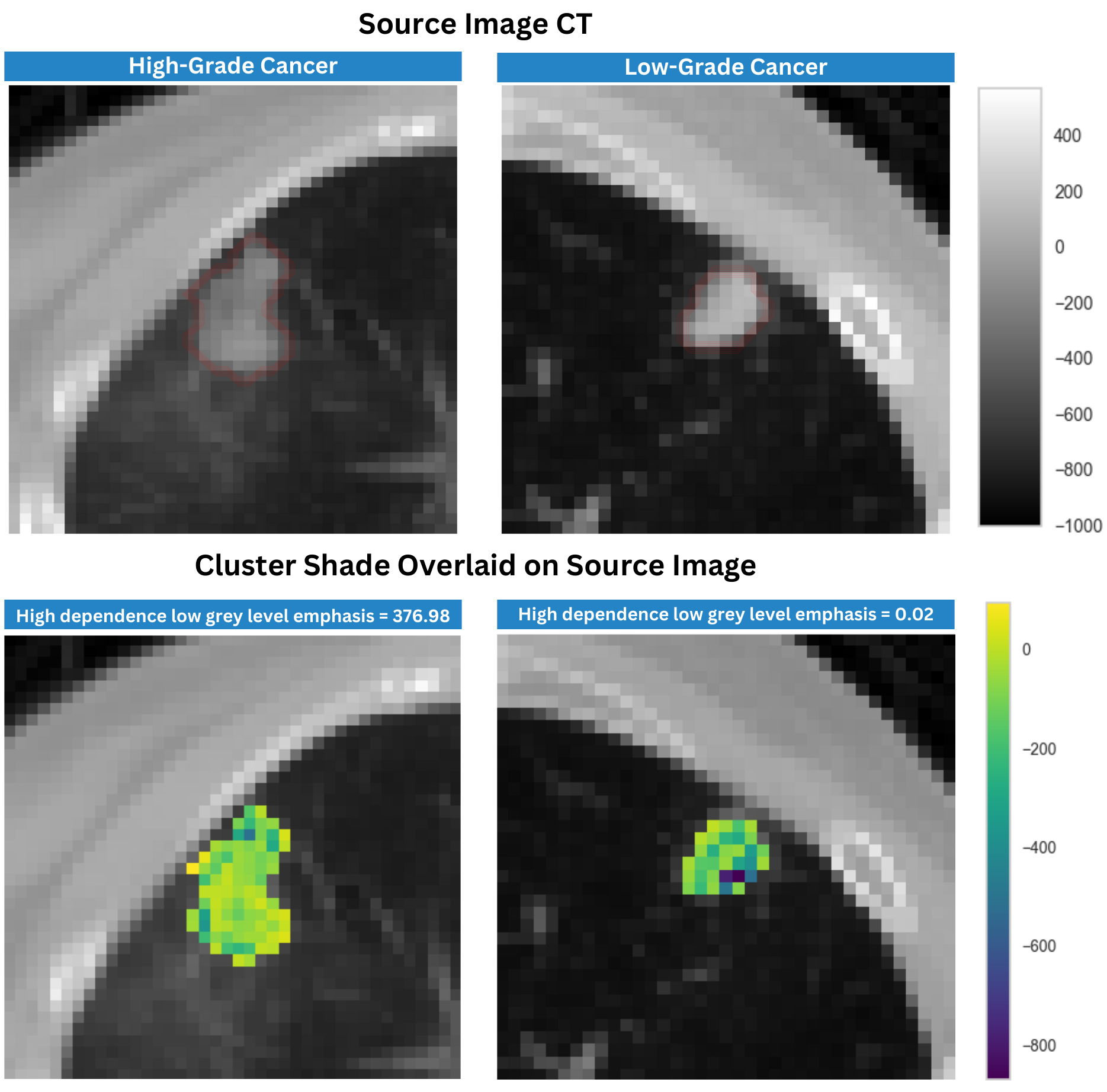}
  \caption{\textbf{CT images illustrating cluster shade textural filtering in non-small cell lung cancer.} Images were selected based on the highest difference in high dependence low gray level emphasis (hdlge) feature (376.98 for high-grade and 0.02 for low-grade cancer), which had the highest importance after training. A contrast difference is visible between low- and high-grade cancers (top). Overlay of the cluster shade revealed intratumoral heterogeneity (bottom).}
  \label{fig:tf}
\end{figure}

Similarly, TF-based models, though not selected as optimal in any cohort, demonstrated high AUC values across various cohorts. For instance, in NSCLC subtype classification, the TF-based model matched the AUC of the selected optimal level (0.76; 95\% CI: 0.72, 0.80), but was not considered optimal due to its higher complexity (detailed results provided in supplementary note 5 of \cite{loutfi_complexity_2024}). Figure \ref{fig:tf} illustrates the application of a textural filter on NSCLC images, selected from patients with the highest difference in the feature with the highest importance.

\subsubsection{In-depth analysis of an optimal complexity level}
\label{sec:in_depth_analysis}
For clear and non-clear RCC classification, the intensity-based level was selected as optimal, and the median intensity had the highest feature importance. To emphasize its impact, images were automatically selected from patients with the highest difference in the median intensity measure and were displayed to visually assess the distinctions between clear and non-clear cell carcinoma based on CECT (See Fig. \ref{fig:analyse_app}.A) (For other cohorts, comparisons are available in supplementary note 4 of \cite{loutfi_complexity_2024}). The clear-cell subtype typically exhibited hypervascularity and greater heterogeneity due to necrotic areas compared to the non-clear cell subtype. Moreover, identifying the optimal level allowed us to refine feature extraction settings, particularly the re-segmentation range \cite{ibsi_manu_1_25}, which directly affects the intensities inside the ROI \cite{ct_range_impact}. This refinement led to a 0.04 gain in AUC from 0.82 to 0.86 (See Fig. \ref{fig:analyse_app}.B).

\begin{figure}
  \centering
  \includegraphics[scale=0.5]{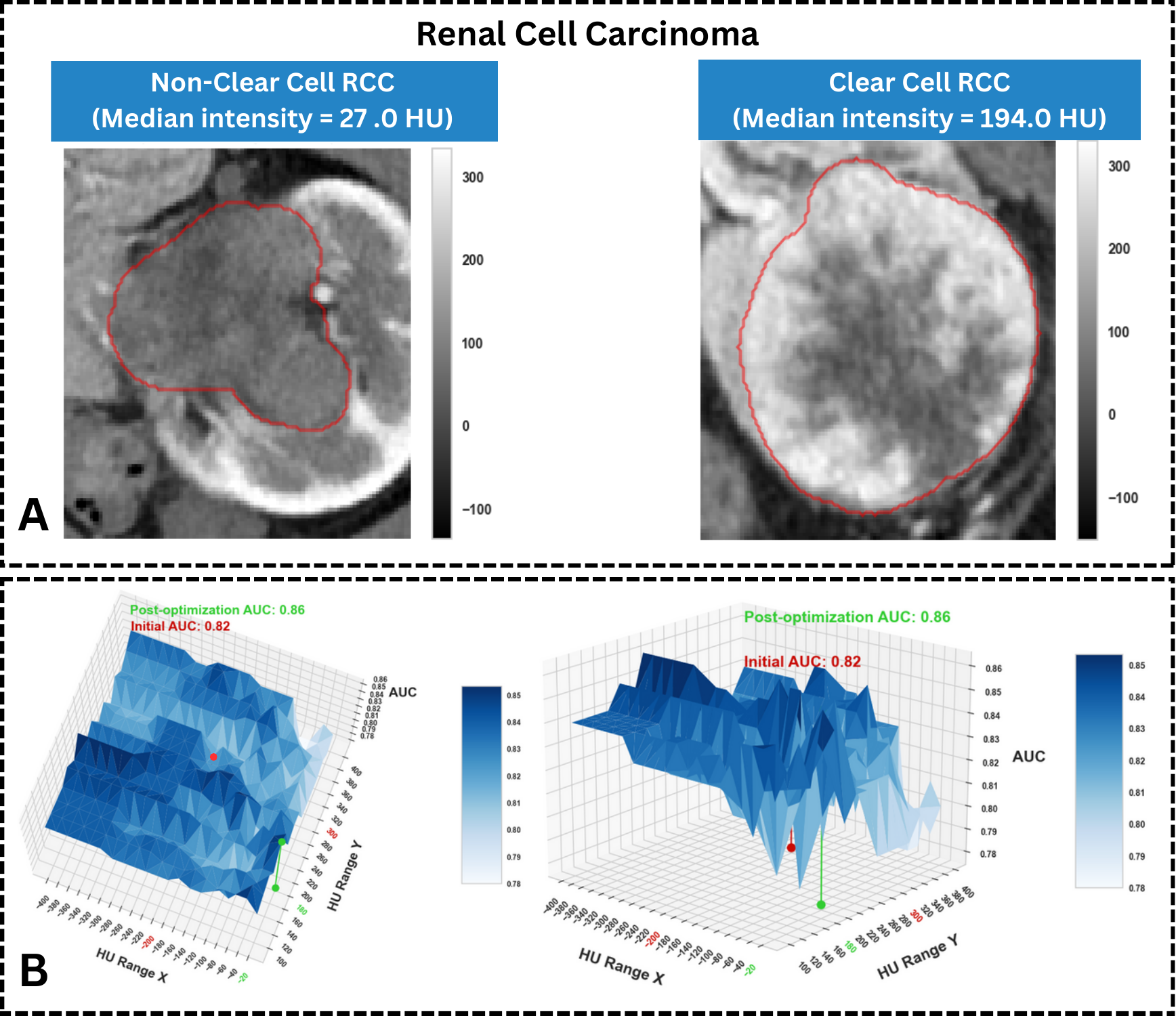}
  \caption{\textbf{Intensity-based analysis of clear cell versus non-clear cell renal cell carcinoma.} (A) Axial contrast-enhanced computed tomography (CECT) comparison of a clear cell versus non-clear cell renal carcinoma with the highest disparity in median intensity. (B) Different views of three-dimensional heatmap illustrating the influence of Hounsfield Units (HU) range $[X,Y]$ on machine learning results in clear cell and non-clear cell renal carcinoma classification. Range limits highlighted in red denote the pre-optimization limits, while those in green signify the updated limits post-optimization.}
  \label{fig:analyse_app}
\end{figure}

\section{Discussion}
\label{sec:discussion}
\textit{MEDiml} combines graphical and programmatic interfaces to radiomics analysis with a complexity-driven approach to model comparison \cite{loutfi_complexity_2024}. The discussion first considers the software as a whole, including its strengths, limitations and future development, then examines the findings and limitations of the five-cohort use-case evaluation.

\subsection{The \textit{MEDiml} software}
\label{sec:discussion_mediml}
\textit{MEDiml} was developed to address two major barriers that continue to limit the adoption of radiomics in clinical research: the technical complexity of radiomics workflows and the lack of synergy between clinical and computational researchers. The platform combines a graphical user interface and a Python package within a unified underlying computational framework, allowing users to construct, execute, and reproduce radiomics workflows through either a visual or programmatic environment. In addition, \textit{MEDiml} introduces the concept of radiomic complexity levels, providing a structured methodology for identifying compact and informative biomarker signatures rather than relying on increasingly large feature sets. Collectively, these features establish \textit{MEDiml} as both a software platform and a methodological framework for radiomics analysis.

The radiomics field has demonstrated considerable potential for clinical decision support; however, challenges related to reproducibility, workflow standardization, interpretability, and feature dimensionality continue to hinder clinical translation \cite{how_to_7, vulnerabilities_8}. \textit{MEDiml} was designed as a response to these challenges. By organizing radiomic features according to their extraction complexity and systematically evaluating predictive performance across complexity levels, the platform promotes the selection of simpler and more interpretable models while maximizing predictive accuracy. This approach differs from many conventional radiomics workflows, which often prioritize the extraction and evaluation of the largest possible feature set. Importantly, compact radiomic signatures are not only easier to interpret but are also better suited for integration with complementary data modalities such as clinical, pathological, genomic, and molecular variables, thereby supporting future multimodal predictive modeling strategies.
Compared with existing radiomics software, \textit{MEDiml} occupies a unique position by combining IBSI-compliant \cite{ibsi_1, ibsi_2} feature extraction, integrated machine learning, local execution, a graphical workflow editor, a Python package, and automatic code generation within a single software ecosystem. While individual tools provide some of these capabilities, \textit{MEDiml} unifies them within a reproducible end-to-end workflow.
The primary strengths of \textit{MEDiml} are its accessibility, reproducibility, and focus on model simplicity. Nevertheless, several limitations remain. First, despite the graphical interface, effective use of the platform still requires familiarity with radiomics concepts and workflows, making comprehensive documentation an important prerequisite for new users. A guided onboarding system and embedded tutorials would help further reduce the learning curve. Second, evaluation of the software has thus far relied primarily on internal collaborators and research partners. Broader adoption and external usability studies are required to identify workflow bottlenecks, missing functionalities, and opportunities for improvement. Finally, the current platform focuses primarily on handcrafted radiomic features and conventional machine learning approaches, and incorporating deep neural networks and multimodal learning frameworks would enable researchers to investigate higher-complexity representations while continuing to compare them against simpler and more interpretable radiomic models. 
Future development of \textit{MEDiml} will focus on expanding the analytical capabilities of the platform while preserving its complexity-driven philosophy \cite{loutfi_complexity_2024}. Planned developments include the integration of deep learning methodologies, support for additional imaging formats, expanded feature selection and modeling strategies, and enhanced tools for multimodal data integration. Such extensions would further strengthen \textit{MEDiml} as a platform for reproducible, explainable, and clinically translatable imaging biomarker research.

\subsection{Use Cases: \textit{MEDiml} Evaluation}
\label{sec:discussion_use_cases}
Overall, in the evaluation of \textit{MEDiml} across the five different oncological datasets, the selection of optimal radiomics complexity levels is aligned with findings from other available studies, indicating the robustness of the complexity-driven approach \cite{loutfi_complexity_2024}. For example, shape features had the highest impact in meningioma grade classification, which corroborates the findings of Zhang \textit{et al.} \cite{zhang_26} who reported a correlation between shape features and brain invasion in meningioma cancer grade prediction. Similarly, texture features were found predictive and sufficient in MRI-based non-clear cell and clear cell RCC classification, corroborating the findings of Wang \textit{et al.} \cite{wang_rcc_27}, who found texture features effective in differentiating three RCC subtypes (clear cell, papillary, and chromophobe) from MR images. Finally, E Linning \textit{et al.} \cite{lungs_linning_28} utilized radiomics for classifying histological subtypes of lung cancer and found intensity features to be the most predictive, indicating tumor heterogeneity. These findings are concordant with our results and suggest our methodology could pave the way for more generalizability across diverse clinical scenarios.

One of the aims in developing the \textit{MEDiml} software around the radiomics complexity level idea was to demonstrate that simpler features often suffice, streamlining radiomics model development, reducing computational demands, improving model interpretability, and potentially increasing generalizability, especially in scenarios where imaging parameters vary significantly. Through this approach, the focus on a singular optimal level can save time and costs by preventing the investigation of irrelevant features, while also laying the groundwork for a more in-depth analysis. This was found to be the case in the clear cell versus non-clear cell RCC classification, where 18,112 features were extracted and tested; however, only a set of 15 intensity-based features was sufficient to obtain the best performance. We then optimized the Hounsfield unit range used in the ROI re-segmentation step, which directly affects the aforementioned features, and successfully improved the AUC from 0.82 to 0.86.

The use-case evaluation \cite{loutfi_complexity_2024} has some limitations. Limited access to large datasets restricted our ability to assess generalizability of the identified optimal complexity levels across different populations, imaging protocols, and clinical settings. Our estimation of optimal complexity levels is susceptible to variations in image processing and feature extraction parameters. All classifications were limited to binary classification problems and other potential radiomics tasks were not explored. Additionally, we did not assess the robustness of features against differences in positioning, acquisition and segmentation \cite{zwanenburg_robustness_29}, potentially leaving the identified optimal levels susceptible to differences in these factors, and our reliance on pre-existing manual segmentations introduces a critical consideration regarding feature robustness and its potential influence on the identified optimal complexity levels. Image segmentation remains a potential bottleneck in radiomics workflows, and future work should ideally incorporate (semi-)automated segmentation methods or explicitly evaluate the robustness of features across different complexity levels to segmentation perturbations to confirm whether the performance of predictive features holds when segmentation variability is minimized or accounted for.

These limitations highlight the necessity of further validation of the complexity-driven approach \cite{loutfi_complexity_2024}, particularly through systematic comparisons with deep radiomics features derived from deep learning or foundation models that could potentially unravel unknown predictive features with high reproducibility potential. Future work should also focus on developing yet more automated strategies capable of estimating the appropriate level of radiomics complexity required for a given clinical task, ideally without the need for human analysis interventions used in the use-case evaluation. Such efforts will contribute to establishing a robust approach for selecting between simpler radiomics or more complex deep learning features.

\section{Conclusion}
\label{sec:conclusion}
This reference manual has presented \textit{MEDiml} as an open-source platform through which clinical researchers and computational scientists can configure and conduct radiomics analyses using a shared computational implementation. Its Python package (\textit{MEDiml-pkg}) and graphical application (\textit{MEDiml-app}) connect image processing, standardized feature extraction, feature set reduction, model training and results analysis. The manual describes how these stages are configured through package classes and graphical modules, together with installation, the software architecture, accelerated filtering and dose maps analysis support.

Code generation allows visually configured pipelines to be inspected and extended as executable Jupyter Notebooks, while statistical comparisons, performance heatmaps and the Explanation Tree (hierarchical feature importance) support the examination of model performance and contributing features. The complexity-driven workflow compares specified feature sets rather than assuming that the largest set is preferable. Finally, as a scientific proof of concept, five oncological use cases illustrate that the selected complexity level depends on the clinical task, with simpler feature groups sufficient in several evaluated settings.

Overall, \textit{MEDiml} provides a documented framework for conducting and interpreting radiomics research, with software, tutorials and support accessible through \href{https://www.mediml.app/}{mediml.app}.

% In conclusion, MEDiml provides a unified and reproducible platform for radiomics research that combines IBSI-compliant \cite{ibsi_1, ibsi_2} feature extraction, integrated machine learning, a no-code graphical interface, and a Python package within a single software ecosystem. Beyond its software contributions, MEDiml introduces the concept of radiomic complexity levels, a methodology for systematically identifying the simplest set of predictive imaging biomarkers required for a given clinical task. The use-case evaluation \cite{loutfi_complexity_2024} across five oncological datasets demonstrated that the most informative radiomic feature categories vary according to the clinical context and that simpler feature representations can often achieve predictive performance comparable to, or better than, more complex alternatives. These findings highlight the importance of prioritizing compact, interpretable, and reproducible radiomic signatures rather than maximizing feature quantity or complexity. Finally, MEDiml aims to bridge the gap between clinical researchers and computational scientists, democratize access to radiomics research, and accelerate the translation of imaging biomarkers into routine clinical practice.

\section*{Acknowledgments}
We thank Guillaume Blain, Sarah Denis, and Andréanne Allaire for their valuable contributions to the development and enhancement of \textit{MEDiml}'s user interface. We also thank Corentin Gauthier, Clarisse Cheng, and Anthonin Falk for their valuable contributions to the development of the \textit{MEDiml} package. Finally, we gratefully acknowledge Martin Lepage for co-supervising a significant portion of this work \cite{loutfi_complexity_2024}.

\section*{Funding information}
Martin Vallières acknowledges the support of Unité de soutien au système de santé apprenant (SSA) Québec.

\section*{Data sharing statement}
At this time, the following datasets are not publicly shared by the hosting institutions: Non-Small Cell Lung Cancer (UCSF), Meningioma (UCSF, PMH), Renal cell carcinoma 1 (Penn, Mayo, HPH, XYSH) and Renal cell carcinoma 2 (CHUS). The Non-Small Cell Lung Cancer data from MAASTRO is available here: \url{https://doi.org/10.7937/K9/TCIA.2015.PF0M9REI} and data from Stanford is available here: \url{https://doi.org/10.7937/K9/TCIA.2017.7hs46erv}. The Low-grade glioma dataset from Huashan is available upon reasonable request here: \url{https://doi.org/10.1038/s41598-017-05848-2}; \url{https://doi.org/10.1007/s00330-016-4653-3}. The Low-grade glioma dataset from TCGA is available here: \url{http://doi.org/10.7937/K9/TCIA.2016.L4LTD3TK}. The Renal cell carcinoma dataset from TCGA is available here: \url{https://doi.org/10.7937/K9/TCIA.2016.V6PBVTDR}.

%Bibliography
\bibliographystyle{unsrt}
\bibliography{references}

\end{document}